\documentclass[12pt,a4paper]{article}

\usepackage{amsmath}
\usepackage{amssymb}
\usepackage{mathrsfs}
\usepackage{graphicx}
\usepackage{enumerate}

\usepackage[pdftex,colorlinks=true,linkcolor=blue,citecolor=blue]{hyperref}

\numberwithin{equation}{section}

\newtheorem{theorem}{Theorem}[section]
\newtheorem{lemma}[theorem]{Lemma}

\newtheorem{definition}[theorem]{Definition}
\newtheorem{example}[theorem]{Example}
\def\proof{\noindent{\em Proof.~}}

\def\eq#1 { \begin{equation} #1 \end{equation} }
\def\eqn#1{ \begin{eqnarray} #1 \end{eqnarray} }
\def\nn { \nonumber }
\def\half{\frac{1}{2}}
\def\cM{{\mathcal{M}}}

\def\cO{{\mathcal{O}}}

\def\cA{{\mathcal{A}}}
\def\cD{{\mathcal{D}}}
\def\cT{{\mathscr{T}}}
\def\cW{\mathcal{W}}
\def\cX{\mathcal{X}}
\def\cR{{\mathcal{R}}}
\def\d{{\partial}}

\def\Reals{\mathbb{R}}
\def\WF{{\rm WF}}
\def\WFA{{\rm WF}_{\rm A}}
\def\C#1{\left\langle #1 \right\rangle}
\def\ket#1{\left| #1 \right\rangle}
\def\bra#1{\left\langle #1 \right|}

\def\cAbulk{{\mathcal{A}_{\rm bulk}}}
\def\cAbndry{{\mathcal{A}_{\rm bndry}}}
\def\D{{\Delta}}
\newif\ifcolorpics
\colorpicstrue
\begin{document}
%%%%%%%%%%%%%%%%%%%%%%%%%%%%%%%%%%%%%%%%%%%%%%%%%%%%%%%%%%%%%%%%%

\title{Boundary-to-bulk maps for AdS causal wedges
and the Reeh-Schlieder property in holography}

\author{Ian A. Morrison\thanks{
    \href{mailto:imorrison@physics.mcgill.ca}
    {imorrison@physics.mcgill.ca}
  }
  \\ \\
  {\it Department of Physics, McGill University, } \\
  {\it Montreal, QC H3A 2T8, Canada }
}

\maketitle

% \begin{center}
%   \vspace{-0.8cm}
%   {\bf Version 9}
% \end{center}

\abstract{
  In order to better understand how AdS holography works for
  sub-regions, we formulate a holographic version of the Reeh-Schlieder 
  theorem for the simple case of an AdS Klein-Gordon field.
  This theorem asserts that the set of states constructed by acting 
  on a suitable vacuum state with boundary observables contained
  within any subset of the boundary is dense in the Hilbert space 
  of the bulk theory.
  To prove this theorem we need two ingredients which are themselves
  of interest. First, we prove a purely bulk version of Reeh-Schlieder
  theorem for an AdS Klein-Gordon field. This theorem relies on the 
  analyticity properties of certain vacuum states.
  Our second ingredient is a boundary-to-bulk map for local observables
  on an AdS causal wedge.
  % This is the sub-region of Anti-de Sitter space whose conformal
  % boundary is a spherical causal development of Minkowski space.
  This mapping is achieved by simple integral kernels
  which construct bulk observables from convolutions with 
  boundary operators.
  % Although these integral kernels are very distributional in character,
  % basic considerations guarantee that their action
  % on boundary correlation functions is well-defined.
  Our analysis improves on previous constructions of AdS boundary-to-bulk
  maps in that it is formulated entirely in Lorentz signature 
  without the need for large analytic continuation of spatial coordinates.
  Both our Reeh-Schlieder theorem and boundary-to-bulk maps
  may be applied to globally well-defined states
  constructed from the usual AdS vacuum as well more
  singular states such as the local vacuum of an AdS causal wedge
  which is singular on the horizon.
}

\newpage

\tableofcontents

%%%%%%%%%%%%%%%%%%%%%%%%%%%%%%%%%%%%%%%%%%%%%%%%%%%%%%%%%%%%%%%%%
\section{Introduction}
\label{sec:intro}
%%%%%%%%%%%%%%%%%%%%%%%%%%%%%%%%%%%%%%%%%%%%%%%%%%%%%%%%%%%%%%%%%

One of the most striking features of quantum physics
is the phenomena of quantum entanglement.
A familiar example of entanglement in quantum mechanics is 
provided by EPR pairs. 
In local quantum field theory (LQFT) there is a sense in which 
entanglement is stronger, or at least more ubiquitous, and 
this is embodied in
the Reeh-Schlieder theorem \cite{Schlieder:1965aa,Haag:1992aa}.
Consider an LQFT on a manifold $\cM$ and consider the set
of local observables $\cA(\cO)$ whose support is contained in a sub-region
$\cO \subset \cM$. The Reeh-Schlieder theorem states that the set
of states generated by acting on a suitable vacuum state with
members of $\cA(\cO)$ is dense upon the Hilbert space.
In essence, in LQFT observables always have long-range correlations
which, even if small, can be exploited; an observer with
limited spacetime but unlimited resources
can explore the entire Hilbert space.

It is natural to ask how this strong notion of quantum entanglement
fits into the AdS/CFT correspondence \cite{Aharony:1999ti,DHoker:2002aw}, 
or more generally gauge/gravity duality \cite{Horowitz:2006ct}.
In general AdS/CFT is a correspondence between
two quantum theories, a $d$-dimensional boundary conformal
field theory (CFT), and $d+1$-dimensional bulk quantum gravity theory. 
When quantum gravity is sufficiently weak the bulk theory becomes 
local and the duality becomes one between two LQFTs.
In this setting one expects the following formulation of
the Reeh-Schlieder theorem: the set of CFT states constructed by
acting on a suitable CFT vacuum with observables supported in any 
sub-region of the boundary provides a dense set of states for the
\emph{bulk} Hilbert space.

In this paper we prove a holographic form of the Reeh-Schlieder
theorem in the simple case of a Klein-Gordon scalar field on a fixed
AdS background. This requires two ingredients, each of which is
of independent interest. Our first ingredient is a purely
bulk formulation of the Reeh-Schlieder theorem. This theorem applies 
to states built atop suitably analytic vacuum states. Examples
of such states include the usual AdS vacuum as well as the natural
``local vacua'' of AdS sub-regions such as the AdS-Rindler vacuum.
In order to make the theorem holographic we define a 
suitable algebra of local boundary observables $\cAbndry$, 
along with the natural algebra of local bulk observables $\cAbulk$.
We then consider the theory restricted to a sub-region 
$\cW$ of AdS known as an AdS causal wedge.
The conformal boundary of an AdS causal wedge is the causal development 
$\cD$ of a compact, spherical region on an equal time hypersurface of 
the boundary. 
Our second ingredient is the construction of a 1-to-1 map for the 
restricted algebras of observables:
$\cAbndry(\cD) \mapsto \cAbulk(\cW)$.\footnote{Strictly speaking the 
boundary algebra $\cAbndry(\cD)$ is larger than the bulk algebra 
$\cAbulk(\cW)$ so the former contains a representation of the 
latter.} Thus any state constructed by applying bulk observables supported
within $\cW$ to a vacuum state may be equivalently constructed 
by applying boundary observables supported within $\cD$. 
The causal wedge has a length scale (the radius of $\cD$) which we may 
make as small as we like. 
In this way we prove the holographic Reeh-Schlieder theorem.

\begin{figure}[!t]
  \centering
  \ifcolorpics
  \includegraphics[width=0.35\textwidth]{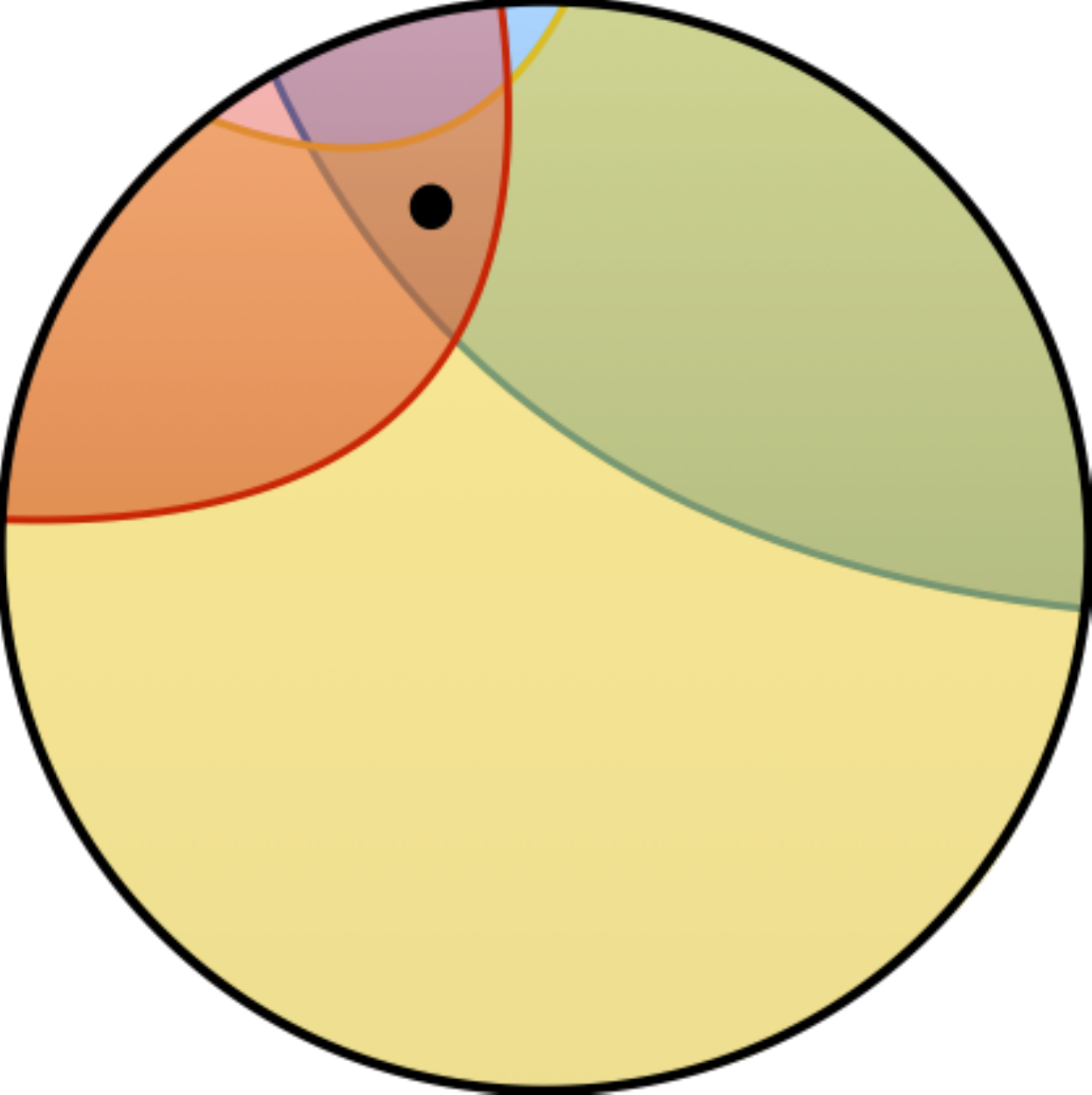}
  \else
  \includegraphics[width=0.35\textwidth]{wedgesbw}
  \fi
  \caption{An an equal-time surface of global AdS. A bulk
    observable is at the black dot. The yellow, red, and blue 
    shaded regions denote AdS
    causal wedges which include the support of the observable.
    The boundaries of these wedges have no mutual intersection.
    The observable may be computed from CFT correlation
    functions on the boundary of any one of these wedges.
  \label{fig:wedges}}
\end{figure}

Aspects of our results have been anticipated in many ways. 
Perhaps the most direct antecedents are discussions of the 
gravity duals of CFT density matrices \cite{Czech:2012bh,Bousso:2012sj}, 
and in particular of the gravity duals of Rindler CFTs
\cite{Emparan:1999gf,Czech:2012be,Parikh:2012kg,Bousso:2012mh}.
Also relevant is the perspective gained from gravity computations
of CFT entanglement entropy \cite{Ryu:2006bv,Hubeny:2007aa,Lewkowycz:2013nqa,
  Faulkner:2013ana} and related quantities (see e.g.,
\cite{Hubeny:2012wa,Casini:2009aa,Hayden:2011ag,Morrison:2012ab,
  Wall:2012uf,Hartman:2013qma,Blanco:2013joa,Kelly:2013aja}).
These investigations provide significant evidence to suggest that
given CFT data on $\cD$ the largest bulk region from which 
we expect to recover ``fine-grained'' information about the bulk theory 
is the associated causal wedge $\cW$. Our boundary-to-bulk map
shows that, at least for this simple theory, this ``fine-grained'' 
information is
in fact the entire bulk observable algebra $\cAbulk(\cW)$.
It is widely believed that data on $\cD$ provides access to
``coarse-grained'' information about the bulk theory beyond $\cW$
(see e.g., \cite{Kraus:2002iv,Fidkowski:2004aa,Hubeny:2006yu,Hubeny:2012ry}).
Our work shows that in this model the course-grained information
accessible is precisely that implied by the Reeh-Schlieder theorem: using 
elements of $\cAbndry(\cD)$ one may well-approximate, but not distinguish, 
any bulk state.

We now describe our process and results in more detail.
In the first part of our study we prove a version of the 
Reeh-Schlieder (RS) theorem for Klein-Gordon fields quantized in the 
Poincar\'e chart of AdS
(the extension to global AdS is straightforward).
As is always the case for the RS theorem, this theorem applies
to states constructed from a suitably analytic vacuum state. 
Our theorem holds for a class of ``locally analytic'' vacuum states 
which are Gaussian states whose Wightman functions satisfy a 
certain analytic wave front set condition. Roughly speaking, these 
vacuum states 
are as analytic as possible within their domain of local analyticity.
The key property enjoyed by these states is a unique notion of
analytic continuation; from this property the RS theorem readily follows.
Since our definition of a locally analytic state is \emph{local} a state
can be locally analytic when restricted to a sub-region of AdS but
fail beyond. 
The usual AdS vacuum is an example of a state which is locally analytic 
on all of the Poincar\'e chart.
Examples of states which are locally analytic only within sub-regions
of AdS include the natural AdS-Rindler vacuum and more generally
the local vacuum of any AdS causal wedge.
These are the natural zero temperature vacuum states defined 
relative to the timelike Killing vector field of the sub-region.
Thus, our RS theorem applies not only to the familiar basis of 
Poincar\'e particle states, but to finite-energy excitations of
these local vacua as well. Essentially any state consistent with
the semi-classical approximation (i.e. a state with well-defined,
finite averaged stress-energy tensor) is included in our
treatment.

In the second part of our study we construct a boundary-to-bulk map 
for local observables on an AdS causal wedge.\footnote{The reader
  should confuse this map with the bulk-to-boundary
  propagator. Our map is a map from the boundary value of 
  the normalizable bulk field to it's bulk profile. There
  are no boundary sources and no non-normalizable part of the
  bulk field.
}
Similar boundary-to-bulk maps for Klein-Gordon fields have been
discussed before, notably in the very nice papers
\cite{Bena:1999jv,Hamilton:2005ju,Hamilton:2006az,Hamilton:2006fh} -- for
generalizations see 
\cite{Lowe:2008ra,Kabat:2011rz,Heemskerk:2012mq,Kabat:2012hp,
  Kabat:2012av}.
While these works represent an excellent initial effort, 
there are some technical improvements we need in order to construct
a successful map for a causal wedge. 
These works employ an integral kernel $K(X|y)$ 
which creates a bulk operator $\Phi(X)$ from it's boundary value 
$\phi(y)$:
\eq{
  \Phi(X) = \int d^d y \sqrt{-\gamma(y)} K(X|y) \phi(y) .
}
Here $X$ is a $d+1$-dimensional bulk coordinate, $y$ is a $d$-dimensional
boundary coordinate, and $\gamma(y)$ is the induced metric on the
conformal boundary.
When considering sub-regions of AdS the integral kernel $K(X|y)$
is a distribution -- in a convenient Fourier representation 
the kernel can diverge exponentially at large momenta -- and thus
it's convolution with correlation functions which are themselves
distributions requires careful analysis.
Previous authors dealt with these divergences by performing
a large analytic continuation in the spatial coordinates of $y$ --
a much larger foray into the complex plane than is implemented
by an $i\epsilon$ prescription -- and this has led some to 
question whether these boundary-to-bulk maps 
are actually well-defined in the physically correct Lorentz
signature \cite{Bousso:2012mh,Leichenauer:2013kaa}.\footnote{Indeed, 
  the analysis of \cite{Bousso:2012mh} and particularly 
  \cite{Leichenauer:2013kaa} shows that in general for less-symmetric,
  asymptotically-AdS spaces the integral kernel $K(X|y)$ is 
  highly distributional. Thus in genereal one cannot side-step 
  the issue via analytic continuations.
%  Perhaps more to the point, analytic continuation obscures the fact
%  that one can construct integral kernels which are compactly supported
%  on the Lorentzian boundary.
}
The more refined analysis we present here shows that despite the
distributional nature of the $K(X|y)$ these boundary-to-bulk maps
converge on states relevant for our RS theorem.

We use the following line of reasoning to show that the boundary-to-bulk 
map for AdS causal wedges converges.
The distributional character of CFT correlation functions is 
constrained by basic aspects of the theory -- ingredients such as
positivity and the non-negativity of the energy spectrum. 
From these considerations we may determine the largest class of
boundary ``test functions'' which have well-defined convolution with 
CFT correlators. This class includes certain distributions in addition
to functions. 
Our main task is to show that the test functions constructed using
the integral kernel $K(X|y)$ appropriate to an AdS causal wedge
are in fact members of this class. Obviously, this analysis is a bit technical,
but it is made rather simple by the use of a tool known
as the $C^\infty$ wave front set of a distribution.
This object characterizes the singularities of a distribution in
an invariant manner.
Lest these tools dismay the reader, we also show through explicit
calculation that the boundary-to-bulk map converges on both the usual
CFT vacuum as well as the local vacuum for a boundary sub-region $\cD$.

This paper proceeds quite simply. In \S\ref{sec:bulk} we set up and prove
the bulk RS theorem. In \S\ref{sec:b2b} we establish the boundary-to-bulk
maps for AdS causal wedges. We close with a discussion
in \S\ref{sec:disc}. Some mathematical
background and lengthy calculations are collected in the Appendices.

%%%%%%%%%%%%%%%%%%%%%%%%%%%%%%%%%%%%%%%%%%%%%%%%%%%%%%%%%%%%%%%%%
\section{The bulk Reeh-Schlieder theorem}
\label{sec:bulk}
%%%%%%%%%%%%%%%%%%%%%%%%%%%%%%%%%%%%%%%%%%%%%%%%%%%%%%%%%%%%%%%%%

%%%%%%%%%%%%%%%%%%%%%%%%%%%%%%%%%%%%%%%%%%%%%%%%%%%%%%%%%%%%%%%%%
\subsection{Bulk theory basics}
\label{sec:bulkBasics}
%%%%%%%%%%%%%%%%%%%%%%%%%%%%%%%%%%%%%%%%%%%%%%%%%%%%%%%%%%%%%%%%%

This section serves to establish our conventions for the bulk 
Klein-Gordon theory. For simplicity we focus on quantization in 
the Poincar\'e chart of $D=d+1$-dimensional AdS (PAdS).
The line element is
\eq{ \label{eq:AdSPoincare}
  ds^2 = \frac{\ell^2}{Z^2}\left( -dT^2 + dZ^2 + d\vec{X}^2 \right) ,
  \quad Z \in (0,+\infty) ,
}
where $\ell$ is the AdS radius and  $Z$ is the AdS ``radial'' direction 
such that $Z=0$ corresponds to the conformal boundary.
The extension of our analysis to global AdS is straightforward.

We consider a real scalar field $\Phi(X)$ obeying the Klein-Gordon
equation
\eq{ \label{eq:KG}
  (\Box - M^2 ) \Phi(X) = 0 .
}
AdS has a timelike conformal boundary the thus
is not globally hyperbolic, so we must impose boundary conditions in
order to have well-posed Cauchy evolution. 
As is well known, solutions to the equation of motion
(\ref{eq:KG}) have two behaviors as $Z \to 0$: 
\eq{
  \Phi(X) \sim Z^\Delta + Z^{d-\Delta} ,
}
where $\Delta = \frac{d}{2} + \sqrt{\frac{d^2}{4}+M^2\ell^2}$ 
such that $M^2\ell^2 = \Delta(\Delta-d)$. We impose the standard boundary
condition that $\Phi(X) = O(Z^\Delta)$ as $Z \to 0$. This
boundary condition is invariant under AdS isometries and 
is sufficient to guarantee unique Cauchy evolution
\cite{Breitenlohner:1982bm,Breitenlohner:1982jf,Ishibashi:2004wx}.

Upon quantization the scalar field becomes an operator and
must be evaluated within a correlation function $\C{\dots}_\Psi$
of some quantum state $\Psi$. Correlation functions 
satisfy the equation of motion
\eq{
  (\Box - M^2)\C{A \Phi(X) B}_\Psi = 0 ,
}
where $A$,$B$ are arbitrary operator insertions, 
as well as the canonical commutation relations
\eq{ \label{eq:CCR}
  \C{ A \left[\Phi(X_1),\Phi(X_2)\right] B}_\Psi
  = i \Delta(X_1,X_2) \C{A B}_\Psi ,
}
where $\Delta(X_1,X_2)$ is the commutator function, a.k.a. the 
advanced-minus-retarded fundamental solution \cite{Friedlander:1975aa},
which is unique given our choice of boundary conditions.
We take as the basic observables of the theory the ``smeared''
operators
\eq{ \label{eq:PhiF}
  \Phi[F] := \int d^D X \sqrt{-g(X)} F(X) \Phi(X) , \quad
  F \in C^\infty(\text{PAdS}) .
}
We define the bulk algebra of local observables $\cAbulk$ to 
be the unital $*$-algebra generated by finite sums of finite
products of the basic elements (\ref{eq:PhiF}).
The restriction of this algebra to a sub-region $\cO \subset \text{PAdS}$
is denoted $\cAbulk(\cO)$.
Quantum states are positive linear functionals on this algebra.

Technically the algebra $\cAbulk$ does not include observables
constructed from composite operators such as $\Phi^2(X)$ or
the all-important stress tensor. Obviously we must have these
observables in order to completely describe bulk physics.
In particular, we regard a finite averaged stress-energy tensor to 
be an additional criteria for a state to be deemed physically 
reasonable (otherwise the semi-classical approximation is invalid).
In this work we consider states
for which composite operators may be constructed via normal ordering.
Thus for these states one may construct directly from $\cAbulk$
a larger ``Wick polynomial algebra'' which includes observables
constructed from composite operators, though we need not present
these details here -- see, e.g., \cite{Hollands:2001fk,Hollands:2002ux,
  Hollands:2004yh,Brunetti:1995rf}.

%%%%%%%%%%%%%%%%%%%%%%%%%%%%%%%%%%%%%%%%%%%%%%%%%%%%%%%%%%%%%%%%%
\subsection{Locally analytic states and the RS theorem}
\label{sec:RS}
%%%%%%%%%%%%%%%%%%%%%%%%%%%%%%%%%%%%%%%%%%%%%%%%%%%%%%%%%%%%%%%%%

For typical field theories on AdS there exists a class of
states which are highly analytic; this can be traced back to the
fact that AdS is a complex analytic manifold.
While a mathematician might call these states sparse in the same
way that analytic functions are sparse on the set of smooth functions,
for most physical questions this set of analytic states provide
the vacuum states of interest. In this section we define a set 
of \emph{locally analytic} states which, roughly speaking, are 
as analytic as the usual AdS vacuum.
This analyticity property is local so a
state may be locally analytic in a sub-region of AdS and fail
to be so elsewhere.
Locally analytic states enjoy a notion of unique analytic 
continuation. This is the property we need to formulate a
Reeh-Schlieder theorem.

For simplicity we restrict attention to quasi-free (a.k.a.~Gaussian)
states. The analyticity properties of a quasi-free state are simply
those of it's Wightman function. A Wightman function is a distribution,
and the analyticity of distributions is considerably more nuanced than 
that of functions. 
The analyticity properties of a distribution are nicely encoded
in a mathematical tool known as the \emph{analytic wave front 
set} ($\WFA$) of a distribution. This is a rather technical
object, so for the moment we provide a colloquial description of 
wave front sets that will likely be sufficient for a first reading. 
We provide a precise definition of the wave front set, along with 
further introduction, in Appendix~\ref{app:WF}.
Let us first describe the $C^\infty$ wave front set ($\WF$)
which we will use later in \S\ref{sec:b2b}.
This object provides a precise characterization of the singularities
of a distribution by listing, for each point in position space at which the
distribution is singular, the directions in a locally-constructed
Fourier space for which the function fails behave as a smooth function.
The wave front set is defined locally and transforms covariantly
under general diffeomorphisms; it is naturally thought of 
as a subset of the cotangent bundle of the manifold.
One may similarly define the analytic wave front set 
which describes the locations and momentum directions in which
a distribution fails to be analytic -- for details see
Appendix~\ref{app:WF}.

Consider the familiar global AdS vacuum $\Omega$.
The Wightman function of $\Omega$ is
\cite{Burgess:1985aa,DHoker:2002aw}
\eqn{ \label{eq:WOmega}
  W_\Omega(X_1,X_2) &:=&
  \C{\Phi(X_1)\Phi(X_2)}_\Omega
  \nn \\
  &=&  \frac{\ell^{(2-D)}\Gamma(\Delta)\xi_{12}^{\Delta}}
  {2^{\Delta+1}\pi^{d/2} \Gamma\left(\Delta-\frac{d}{2}+1\right)} 
  {}_2 F_1\hspace{-4pt}\left[ \frac{\Delta}{2}, \frac{\Delta+1}{2} ; 
    \Delta-\frac{d}{2}+1 ; \xi_{12}^2 \right] .
}
Here ${}_2F_1$ is the Gauss hypergeometric function,
$\xi_{12} = (1 + u_{12})^{-1} + i\epsilon s_{12}$, 
$u_{12}$ is the $SO(d,2)$-invariant chordal distance between $X_1$ 
and $X_2$,\footnote{
  % In terms of the embedding space (\ref{eq:AdSembedding})
  % the chordal distance is
  % \eq{
  %   u_{12} = \frac{1}{2\ell^{2}} ||\cX_1 - \cX_2||^2 .
  % }
  In Poincar\'e coordinates (\ref{eq:AdSPoincare})
  \eq{
    \xi_{12} = \frac{2 Z_1 Z_2}{Z_1^2 + Z_2^2 + (x_1 - x_2)^2} ,
  }
  where we use the notation $X = (Z,x)$.
}
and $s_{12} = +(-)1$ for $X_1$ in the causal future (past) of $X_2$ and
equal to $0$ else. This Wightman function is analytic away from 
$\xi_{12} = +(-) 1$ corresponding to null (antipodal) separation.
More precise information is given by it's analytic wave front set:
\eq{\label{eq:WFOmega}
  \WFA(W_\Omega) = \left\{
    (x_1,k_1;x_2,k_2) \in (T^*(\text{PAdS}))^2 \setminus \{0 \} \; \Big| \;
    \xi_{12} = \pm 1,\; k_1 \in V^+, \; k_1 \sim - k_2
    \right\} .
}
Following notation employed in Appendix~\ref{app:WF}, $T^*$ denotes
the tangent space, $\{0\}$ the zero section, $V^+$ the closed
forward lightcone,\footnote{The closed forward (backward) cones are 
  defined to be
  $V^{+(-)} := \{ k \in \Reals^{d-1,1} \; | \; k^2 \le 0,\; + (-) k^0 > 0 \}$.
  We use Lorentz signature for Fourier space and the tangent space.
} and $k_1 \sim k_2$ denotes an equality after parallel
transporting $k_1$ to $x_2$.
Essentially (\ref{eq:WFOmega}) states that the singularities of 
$W_\Omega$ are of locally positive frequency, much like the 
coincident singularity of the Poincar\'e-invariant vacuum of Minkowski 
QFT.

One could also consider states which have the same analyticity
properties of the AdS vacuum in a sub-region of PAdS but fail to be
analytic everywhere. Examples of such states
include the ``local vacua'' of AdS causal wedges and the AdS-Rindler
wedge. The Wightman function of such a state is as analytic as $\Omega$ 
within it's defining sub-region, but has additional singularities on 
the boundary horizon as well as beyond the region. With these states in 
mind we define:
\begin{definition}
  A locally analytic state $\Psi$ on $\cAbulk(\cO)$ is a quasi-free
  state whose Wightman function $W_\Psi(X_1,X_2)$ 
  satisfies the analytic wave front set condition
  \eq{ \label{eq:WFLA} 
    \WFA\left(W_\Psi\right) \subseteq \left\{
      (x_1,k_1;x_2,k_2) \in (T^*(\cO))^2 \setminus \{0 \} \; \Big| \;
      \xi_{12} = \pm 1,\; k_1 \in V^+, \; k_1 \sim - k_2
    \right\} .
}
\end{definition}
The AdS vacuum $\Omega$ is locally analytic on all of PAdS;
we will verify later that natural vacuum of an AdS causal wedge
is locally analytic on the wedge.

\begin{figure}
  \centering
  \ifcolorpics
  \includegraphics[width=0.25\textwidth]{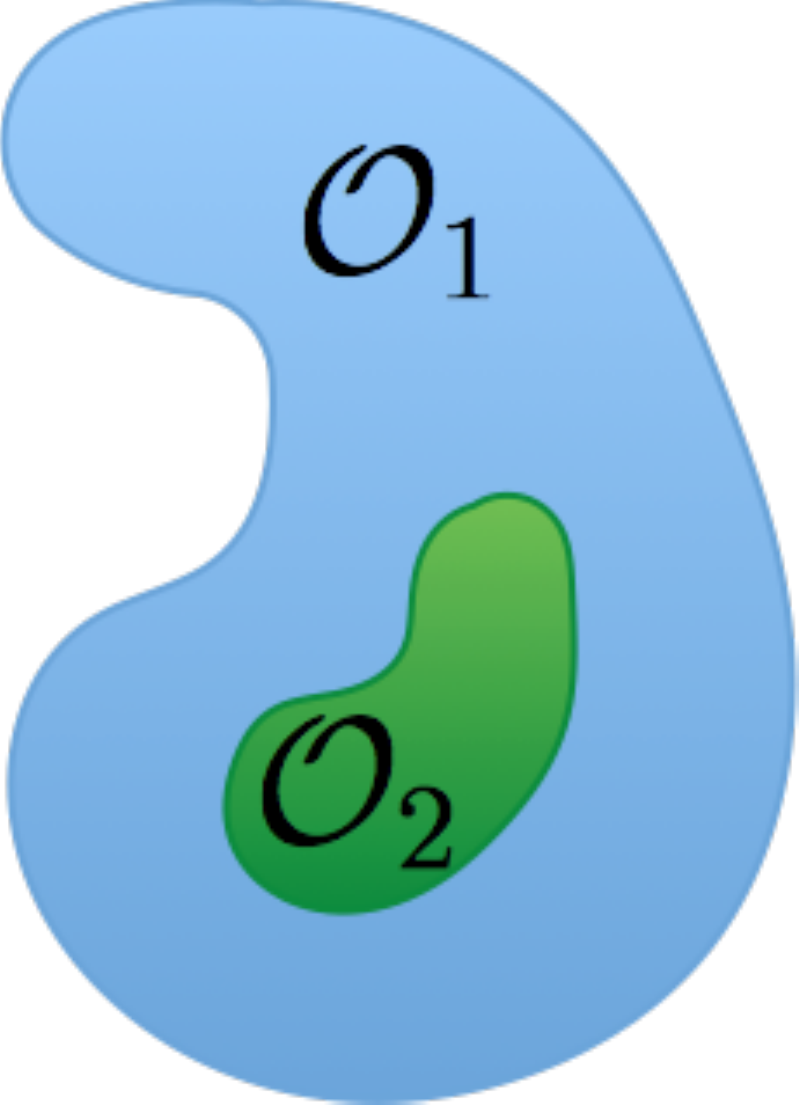}%
  \hspace{1cm}%
  \includegraphics[width=0.4\textwidth]{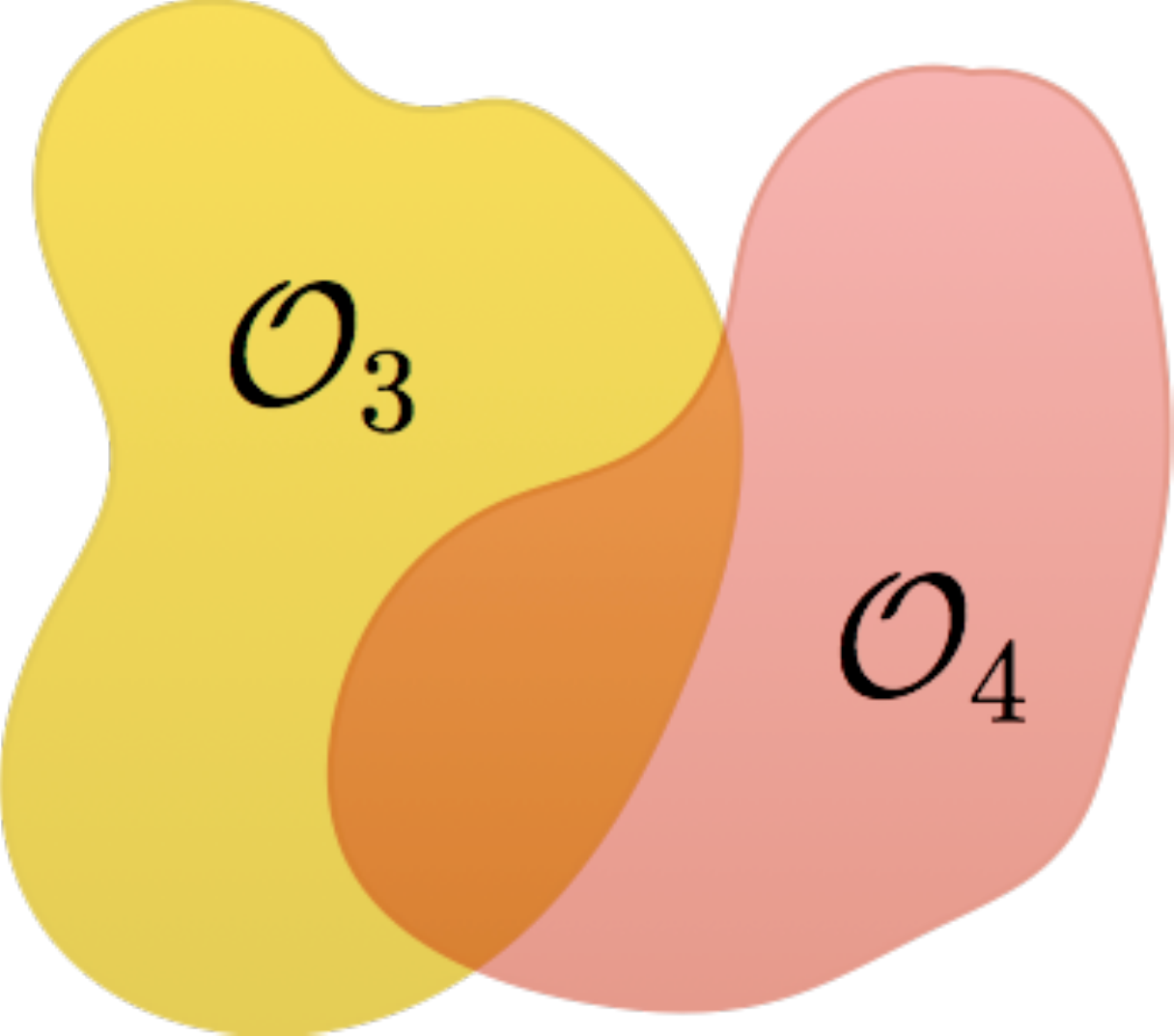}\\
  \else
  \includegraphics[width=0.25\textwidth]{domains1bw}%
  \hspace{1cm}%
  \includegraphics[width=0.4\textwidth]{domains2bw}\\
  \fi
  \caption{Depictions of the domains of analyticity referenced in the 
    text.\label{fig:domains}}
\end{figure}

Locally analytic states enjoy a notion of unique analytic continuation.
This follows from a very general form of the ``edge of the wedge theorem''
\cite{Streater:1989vi} familiar from Minkowski space QFT reformulated
here in terms of analytic wave front sets:
\begin{theorem}[Edge of the wedge theorem (Proposition 5.3 of 
  \cite{Strohmaier:2002aa})] \label{thm:EOTW}
  Let $\cM$ \\
  be a real analytic connected manifold and $u$
  a distribution (in the distribution space dual to that of smooth functions)
  with the property that
  \eq{ \label{eq:EOTW}
    \WFA(u) \cap - \WFA(u) = \emptyset .
  }
  Then for each non-void open subset $\cO \subset \cM$ if the restriction of
  $u$ to $\cO$ vanishes then $u = 0$. 
\end{theorem}
The following simple lemmas follow immediately:
\begin{lemma} \label{lem:ac1}
  Let $\Psi$ be locally analytic on $\cO_1$.
  If a $2n$-pt function of $\Psi$ vanishes when restricted
  to an open set $\cO_2 \subset \cO_1$ then $\Psi = 0$ on $\cO_1$.
\end{lemma}
\proof The state $\Psi$ is quasi-free so if a 2n-pt function 
vanishes on $\cO_2$ it implies $W_\Psi = 0$ on $\cO_2$. 
$W_\Psi$ satisfies the criteria (\ref{eq:EOTW}) so if it vanishes
on $\cO_2$ it vanishes on $\cO_1$. Thus $\Psi = 0$ on $\cO_1$. $\blacksquare$
\begin{lemma} \label{lem:ac2}
  Let $\Psi_{3/4}$ be locally analytic on $\cO_{3/4}$,
  and let $\cO_3 \cap \cO_4$ be non-void.
  If $W_{\Psi_3} - W_{\Psi_4} = 0$
  on an open set of $\cO_3 \cap \cO_4$ then 
  $\Psi_3 = \Psi_4$ on $\cO_3 \cap \cO_4$ and $\Psi_{3/4}$ provides the
  unique locally analytic extension of $\Psi_{4/3}$ on $\cO_{3/4}$.
\end{lemma}
\proof This follows immediately from the fact that $W_{\Psi_{3/4}}$ each
satisfy (\ref{eq:EOTW}). $\blacksquare$
\vspace{8pt}\\
These lemmas show how the locally analytic extension of a state is
uniquely defined, but they do not prove the existence of such an extension.
Indeed, as the example of AdS causal wedge vacua show, in general there 
is no such extension.

Let us turn the
discussion to the more general class of states constructed by acting
on a locally analytic state with members of $\cAbulk$.
It is easy to show that such states are no more singular than
locally analytic states, but are not necessarily as
analytic. The following lemma follows immediately
from the properties of wave front sets:
\begin{lemma}
  Let $\Psi$ be locally analytic on $\cO_1$ and let
  $\cO_2 \subset \cO_1$ be a sub-region such that it's causal
  complement $\overline{\cO}_2$ on $\cO_1$ is non-void.
  Then the state $\Psi_2 = A_2 \Psi$ with $A_2 \in \cAbulk(\cO_2)$
  is locally analytic on at least $\overline{\cO}_2$.
\end{lemma}
% \begin{lemma}
%   Let $\Psi$ be locally analytic on $\cO_1$ and let 
%   $\Psi_1 = A_1 \Psi$ with $A_1 \in \cAbulk(\cO_1)$.
%   The Wightman function of $\Psi$ restricted to $\cO_1$ 
%   has the $C^\infty$ wave front set
%   $\WF(W_\Psi) \subseteq \WFA(W_{\Psi})$, the latter of 
%   which is given by (\ref{eq:WFLA}).
% \end{lemma}
Despite being less analytic, states generated from locally
analytic states enjoy a weaker 
property reminiscent of analyticity which is embodied in the 
Reeh-Schlieder theorem:
\begin{theorem}[Reeh-Schlieder theorem] \label{thm:RS}
  Let $\Psi$ be locally analytic on $\cO_1$ and let
  $\cO_2 \subset \cO_1$.
  The linear spans of the sets
  \eq{
    S_1 = \{ A_1 \Psi \; | \; A_1 \in \cAbulk(\cO_1) \}, \quad
    S_2 = \{ A_2 \Psi \; | \; A_2 \in \cAbulk(\cO_2) \}
  }
  are equal.
\end{theorem}
\proof Let $\Psi_1 \in S_1$ and let $C = \Phi(X_1)\dots\Phi(X_{2n})$
be any finite string of operators with arguments restricted to
$\cO_2$. Consider
\eq{
  F = \bra{\Psi_1} C \ket{\Psi} = \C{A_1^* C }_\Psi  .
}
If $F$ is non-zero for all $A_1 \in \cAbulk(\cO_1)$ then the assertion 
of the theorem holds. If $F=0$ for some $A_1$ then it follows from 
lemma~\ref{lem:ac1} that $\Psi = 0$ on $\cO_1$ and the theorem 
holds. $\blacksquare$
\vspace{8pt}\\
\noindent To understand the implications of this theorem 
consider when $\Psi$ is
the AdS vacuum $\Omega$ and $\cO_1$ is PAdS. Then the set $S_1$ is
the set of finite-energy excitations of $\Omega$, and $S_2$ is the
set of states which may be constructed using observables contained
in $\cO_2$. The RS theorem states 
that there is no state in $S_1$ orthogonal to the set $S_2$. 
If we consider as the Hilbert space of the bulk theory the usual
one for which $\Omega$ is cyclic then it follows that the set
$S_2$ is dense on this Hilbert space. Thus any state in this
Hilbert space may be approximated
with arbitrary precision by a state on $S_2$.
In other words, by judicious application of operators in $\cO_2$
to $\Omega$ one may construct a global state which well-approximates 
any finite-energy state everywhere on PAdS. 
An important corollary is:
\begin{theorem} \label{thm:RSc}
  Consider the same configuration as theorem~\ref{thm:RS}, and
  in addition let $\overline{\cO}_2$ be the non-void causal complement
  of $\cO_2$ on $\cO_1$.
  If $A_2 \in \cAbulk(\cO_2)$ annihilates any state in $S_1$ then
  it annihilates all states in $S_1$. 
\end{theorem}
\proof From theorem~\ref{thm:RS} it follows that the set
$\overline{S}_2 = \{ B_2 \Psi \; | \; 
B_2 \in \cAbulk(\overline{\cO}_2) \}$
spans $S_1$. All such $B_2$ commute with $A_2$. Thus if $A_2\Psi = 0$ 
then $A_2 B_2 \Psi = 0$ and so $A_2$ annihilates $S_1$. 
The conclusion is unchanged if we exchange $\Psi$ for any member $A_1 \Psi
 \in S_1$. $\blacksquare$
\vspace{8pt}\\
This corollary makes precise the difference between 
a dense set of states and the entire Hilbert space.
If we think again of the case where $\Psi = \Omega$ and $\cO_1 = \text{PAdS}$
then the theorem states that an observer confined
to $\cO_2$ cannot construct a set of exact annihilation operators
for the span $S_1$ and thus the $\cO_2$ observer cannot exactly
determine the quantum state. For further interpretation of the
RS theorem and related corollaries see, e.g., 
\cite{Haag:1992aa,Streater:1989vi}.

The interpretations we have just given can also be applied to 
cases where $\cO_1$ is a sub-region of PAdS and $\Psi$ is any
locally analytic state on $\cO_1$.

%%%%%%%%%%%%%%%%%%%%%%%%%%%%%%%%%%%%%%%%%%%%%%%%%%%%%%%%%%%%%%%%%
\section{Boundary-to-bulk maps}
\label{sec:b2b}
%%%%%%%%%%%%%%%%%%%%%%%%%%%%%%%%%%%%%%%%%%%%%%%%%%%%%%%%%%%%%%%%%

In this section we change gears and describe the boundary-to-bulk
map for local observables on an AdS causal wedge. After establishing
some technical details in \S\ref{sec:boundary} we review the construction
of such a map for the Poincar\'e chart in \S\ref{sec:Poincare}. We introduce
AdS causal wedges in \S\ref{sec:wedge}. Finally in \S\ref{sec:AdSR} we
examine in detail the boundary-to-bulk map for the AdS-Rindler wedge.

%%%%%%%%%%%%%%%%%%%%%%%%%%%%%%%%%%%%%%%%%%%%%%%%%%%%%%%%%%%%%%%%%
\subsection{Boundary theory basics}
\label{sec:boundary}
%%%%%%%%%%%%%%%%%%%%%%%%%%%%%%%%%%%%%%%%%%%%%%%%%%%%%%%%%%%%%%%%%

In this section we review the construction of the boundary theory 
from the bulk theory. We know this is familiar territory, but we
will need some rather fine details for our later analysis so we
might as well state everything clearly now.

The conformal boundary of the $d+1$-dimensional Poincar\'e chart is 
Minkowski space $\Reals^{d-1,1}$.
Following the standard AdS/CFT prescription we define the boundary
operator $\phi(x)$ via the limit
\eq{ \label{eq:Obndry}
  \phi(x) := \lim_{Z\to 0} Z^\Delta \Phi(X) .
}
We use the notation $X = (Z,x)$ with $x$ a $d$-dimensional
coordinate. AdS transformations in the bulk act as conformal transformations 
on the conformal boundary, and under such actions 
$\phi(x)$ transforms as a conformal field of weight $\Delta$.
Thus $\phi(x)$ constructed in this way may be used to define a 
CFT on $d$-dimensional Minkowski space. This CFT what we refer to as
``the CFT'' or ``the boundary theory.''

We say that every bulk state $\Psi$ induces 
a boundary state $\psi$; so for instance, the AdS-invariant bulk 
state $\Omega$ induces a conformally-invariant boundary state $\omega$.
Obviously, the boundary states induced by 
quasi-free states are also quasi-free. We restrict attention to boundary
states induced by bulk states constructed from locally analytic
states within their domain of analyticity. The 2-pt functions of
such boundary states have reasonable singularity structure.
In fact, they satisfy the so-called ``microlocal spectrum condition''
\cite{Brunetti:1995rf}. For the case at hand this condition states that
the 2-pt function $W_\psi(x_1,x_2)$ of a state $\psi$ has a $C^\infty$
wave front set contained in 
\eq{ \label{eq:muSC}
  \WF(W_\psi) \subseteq \left\{
    (x_1,k_1;x_2,k_2) \in (T^*\Reals^{d-1,1})^2 \setminus \{ 0 \} \; \Big| \;
    |x_1  - x_2|^2 = 0,\; k_1 \in V^+, \;
    k_1 \sim - k_2
    \right\} .
}
In plain words this says that $W_\psi(x_1,x_2)$ is singular at most when
$x_1$ and $x_2$ are null-separated, and that the singularities are
of locally-positive frequency. One may regard (\ref{eq:muSC}) as a very
basic statement about the OPE structure of the theory. 
% It is very easy to see that the usual Minkowski vacuum state $\omega$ 
% satisfies (\ref{eq:muSC}): the 2-pt function of $\omega$ is
% \eqn{ \label{eq:omegaonMx}
%   W_\omega(x_1,x_2) &=&  \C{\phi(x_1)\phi(x_2)}_\omega  \nn \\
%   &\propto& \frac{1}{|x_1 - x_2|^{2\Delta}} \nn \\
%   &\propto& 
%   \int \frac{d^d k}{(2\pi)^d} e^{ik \cdot (x_1-x_2)}
%   \frac{1}{|\omega|}(\omega^2 - k^2)^{2\Delta-d/2} \theta(\omega > |k|) .
% }

For our purposes we define the algebra of local boundary
observables $\cAbndry$ to be the unital $*$-algebra composed of finite sums 
of finite products of the smeared boundary field
\eq{ \label{eq:phif}
  \phi[f] = \int d^dx \sqrt{-\gamma(x)} f(x) \phi(x) , \quad f \in \cT ,
}
where $\gamma(-x)$ is the determinant of the induced metric on the 
boundary and $\cT$ is a suitable class of test functions.
In textbook introductions to Minkowski QFT the class of test functions
is usually taken to be smooth functions, and there are many good physical
and mathematical reasons for this choice \cite{Streater:1989vi,Haag:1992aa}.
For the purposes of holography it turns out that we need a different, strictly
larger class of test functions whose precise definition is somewhat technical.
The issue here is that we desire an algebra of boundary observables which
is large enough to contain compactly-supported representations of 
bulk observables (this will be made more clear in later sections.)
This requires that we consider as ``test functions'' not only 
smooth functions but also certain distributions.
Since correlation functions of $\phi(x)$ are themselves distributions, 
the observables (\ref{eq:phif}) involve the point-wise product of 
distributions and care must be taken to assure that these objects are 
well-defined.

As described in Appendix~\ref{app:WF}, wave front sets allow us
to provide a precise criteria for when the point-wise product of
two distributions is unique. Provided that the boundary
theory 2-pt functions satisfy (\ref{eq:muSC}), the largest class
of ``test functions'' $\cT$ which yield well-defined convolutions 
$\phi[f]$ is the set of distributions on $\Reals^{d-1,1}$ whose 
members $f$ satisfy the wave front set condition
\eq{ \label{eq:cT}
  \WF(f) \subseteq \left\{
    (x,k) \in T^*\Reals^{d-1,1} \setminus \{ 0 \} \; \Big|
    \; k^2 > 0
  \right\} .
}
In plain words this says that $f$ is a distribution whose
local Fourier transform behaves like that of a smooth function 
in timelike and null momentum directions. Thus the convolution
$\phi[f]$ contains no overlapping singular directions and is
unambiguous.

%%%%%%%%%%%%%%%%%%%%%%%%%%%%%%%%%%%%%%%%%%%%%%%%%%%%%%%%%%%%%%%%%
\subsection{The case of the Poincar\'e chart}
\label{sec:Poincare}
%%%%%%%%%%%%%%%%%%%%%%%%%%%%%%%%%%%%%%%%%%%%%%%%%%%%%%%%%%%%%%%%%

We now review the construction of the boundary-to-bulk
map for local observables on Poincar\'e AdS following closely
\cite{Bena:1999jv,Hamilton:2005ju,Hamilton:2006az,Hamilton:2006fh}
-- for some complementary remarks on this process see 
\cite{Heemskerk:2012mn}.
Our goal is to compute local bulk observables in a bulk state 
$\Psi$ given only it's boundary counterpart $\psi$.
In this section we assume we have access to the 2-pt function of 
$\psi$ everywhere on the Minkowski boundary, i.e. we have\footnote{
  States defined on the entire $\Reals^{d-1,1}$ boundary have
  Wightman functions whose support in momentum space is limited to 
  $|k^0| \ge |\vec{k}|$. This follows from the fact that spectrum of 
  the boundary translation operator $P_\mu$ is contained in the closed 
  forward lightcone $V^+$. This is mimicked in the Poincar\'e bulk where
  the restriction $|k^0| \ge |\vec{k}|$ also follows from the non-negativity
  of the energy spectrum.
}
\eq{ \label{eq:2ptbndry}
  \C{\phi(x_1)\phi(x_2)}_\psi = \int \frac{d^d k_1}{(2\pi)^d} 
  \frac{d^d k_2}{(2\pi)^d} \theta(-k_1^2)\theta(-k_2^2)
  e^{i k_1 \cdot x_1} e^{-i k_2 \cdot x_2} \widehat{W}_\psi(k_1,k_2) .
}
A very intuitive way to proceed is as follows. 
Recall that in the bulk there exists a complete set of mode solutions
to the Klein-Gordon equation. We may take these modes to have the form 
$e^{ik\cdot x}V_k(Z)$, where the function $V_k(Z)$ which contains the 
$Z$-dependence and may be 
taken to be real and satisfy $V_k = Z^\Delta(1 + O(Z^2))$ as $Z \to 0$. 
This asymptotic behavior is as prescribed
by our boundary conditions; the normalization is a convenient choice.
Using these bulk Klein-Gordon mode functions we
may extend the 2-pt function of $\psi$ into the bulk simply by
``dressing'' each Fourier mode in (\ref{eq:2ptbndry})
with the appropriate $V_k(Z)$:
\eq{ \label{eq:2ptbulk}
  \C{\Phi(x_1)\Phi(x_2)}_\Psi = \int \frac{d^d k_1}{(2\pi)^d} 
  \frac{d^d k_2}{(2\pi)^d} \theta(-k_1^2)\theta(-k_2^2)
  e^{i k_1 \cdot x_1} e^{-i k_2 \cdot x_2}
  \widehat{W}_\psi(k_1,k_2)
  V_{k_1}(z_1) V_{k_2}(z_2) .
}
By construction this bulk 2-pt function satisfies the equation
of motion and limits to the boundary 2-pt function of $\psi$. If the boundary
2-pt function is positive then so too is the bulk 2-pt function.
Because we assume that $\psi$ satisfies the microlocal spectrum 
condition (\ref{eq:muSC}), it follows that the momentum integrals
in (\ref{eq:2ptbulk}) converge for spacelike-separated bulk points;
for points with timelike or null separation further analysis is required.

The process of computing a bulk observation in $\Psi$ may be regarded
within the boundary theory as an observation of $\psi$. This 
perspective has some technical advantages. 
In this way of looking at things the goal is to show
that for each element $\Phi[F] \in \cAbulk$ there exists
a representative element $\phi[f_F] \in \cAbndry$. Taken
together the set $\{ \phi[f_F] \} \subset \cAbndry$
forms a representation of the bulk algebra of observables.
Alternatively, we can say that there exists a 1-to-1 map
$\cAbndry \mapsto \cAbulk$ which effectively constructs
a bulk state from it's boundary value: $\psi \mapsto \Psi$.
This map is provided by the 
integral kernel\footnote{As has been discussed by previous authors, the 
  representation of the kernel $K(X|y)$ is not unique. This is 
  because the CFT correlation functions on which it acts do not
  have support for spacelike momenta, so one is free to add to
  $K(X|y)$ any object whose Fourier transform is supported only 
  at spacelike momenta.
  Of course the map $\cAbndry\mapsto\cAbulk$ is unaltered
  by such additions.
} \cite{Hamilton:2005ju,Hamilton:2006az,
  Hamilton:2006fh}
\eq{ \label{eq:KPoincare}
  K(X | y) = 
  \int \frac{d^{d}k}{(2\pi)^d} \theta(-k^2)
  e^{i k \cdot (x-y)} V_k(Z) .
}
This object takes one bulk argument $X$ and one boundary argument $y$.
For each bulk test function $F(X)$ we associate a boundary test 
function $f_F(y)$ via
\eq{ \label{eq:f_F}
  f_F(y) := \int d^DX \sqrt{-g(X)} F(X) K(X|y) , \quad
  F \in C^\infty(\text{PAdS}) .
}
By construction the kernel satisfies the bulk equation of motion
with respect to $X$
and is also consistent with the boundary conditions imposed on the
bulk theory. Note that the kernel is a distribution and is not
smooth. For fixed $Z$ the bulk Klein-Gordon mode functions $V_k(Z)$
decay like an inverse power of $|k|$ as $|k| \to \infty$.
What is important, however, is not the behavior of $K(X|y)$ but
that of $f_F(y)$. Because $F(X)$ is smooth a smooth function of $X$
it follows that the $f_F(y)$ defined in (\ref{eq:f_F}) is a smooth
function of $y$,
so indeed $f_F \in \cT$.
The test functions $f_F(y)$ which extract bulk
observables from the boundary theory are essentially as nice as 
can be and any globally well-defined Minkowski CFT state yields
finite correlators for the set $\{\phi[f_F]\}$.

% We note that by focusing on observables rather than the field operators
% themselves our process better mimics the operations 
% available to a boundary observer (which in this case is eternal). 
% Although we do not consider such cases here, there is also an
% obvious example to focusing on observables when the the boundary or
% bulk theory contains local gauge freedom.
% In this case dealing directly with field operators can require tackling
% various gauge condition subtleties.

%%%%%%%%%%%%%%%%%%%%%%%%%%%%%%%%%%%%%%%%%%%%%%%%%%%%%%%%%%%%%%%%%
\subsection{AdS causal wedges}
\label{sec:wedge}
%%%%%%%%%%%%%%%%%%%%%%%%%%%%%%%%%%%%%%%%%%%%%%%%%%%%%%%%%%%%%%%%%

\begin{figure}
  \centering
  \ifcolorpics
  \includegraphics[width=0.2\textwidth]{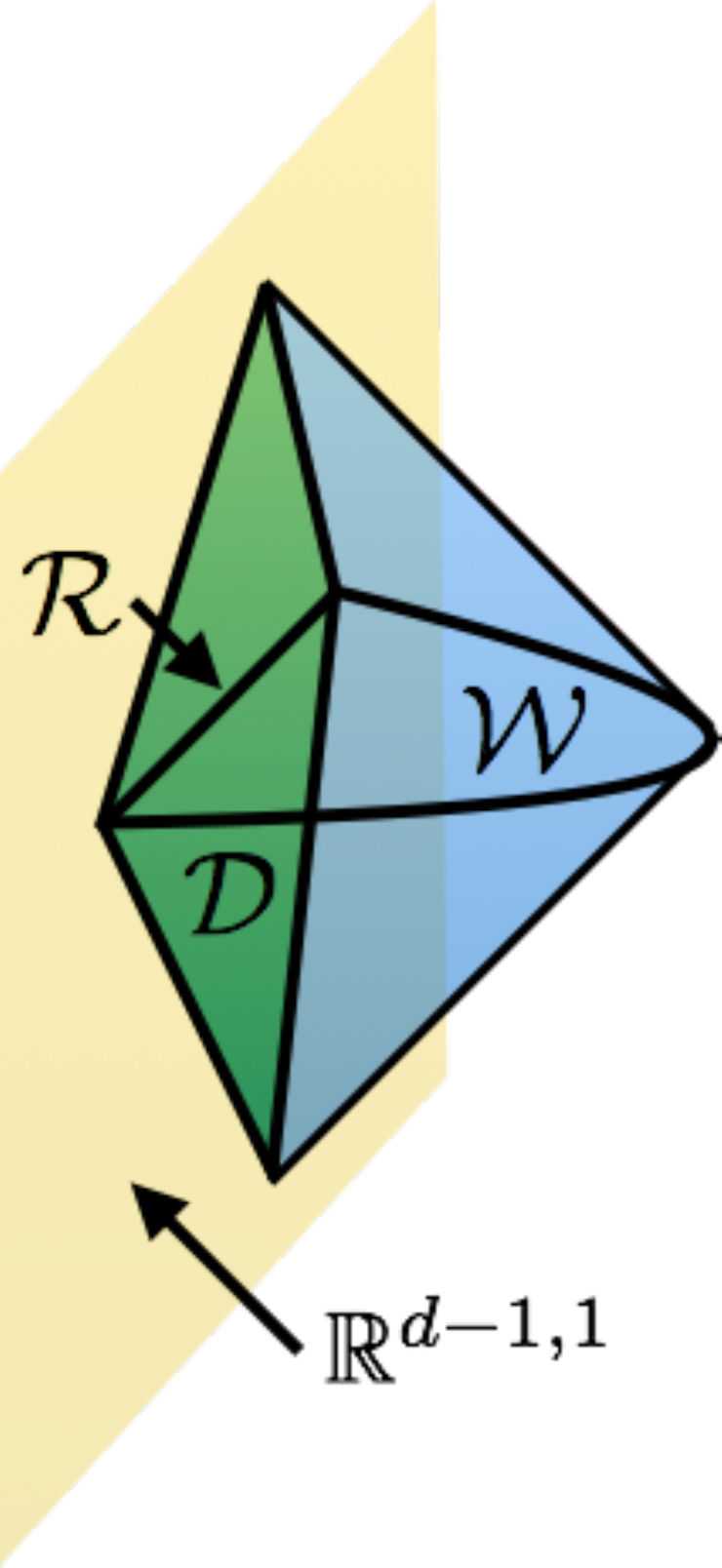}%
  \hspace{3cm}
  \includegraphics[width=0.2\textwidth]{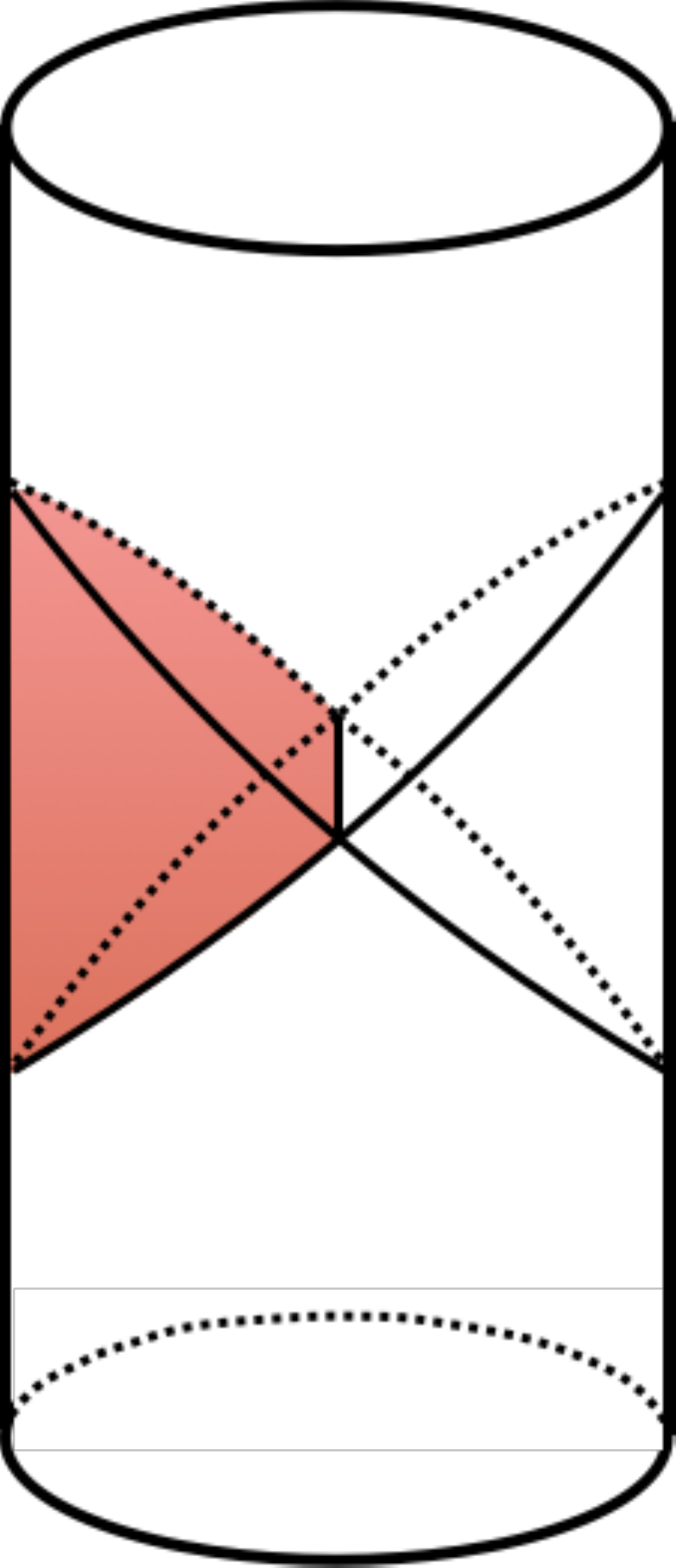}
  \else
  \includegraphics[width=0.2\textwidth]{wedge3d3bw}%
  \hspace{3cm}
  \includegraphics[width=0.2\textwidth]{rindlerWedgebw}
  \fi
  \caption{Left: an AdS causal wedge. Right: global AdS with
    the AdS-Rindler wedge highlighted.
    \label{fig:wedge3d}}
\end{figure}

Next we would like to construct a boundary-to-bulk map for a sub-region
of AdS which is smaller than the Poincar\'e chart and likewise employs
a smaller region of the boundary than the entire Minkowski space.
As has been discussed by many before (see e.g., \cite{Czech:2012bh,
  Bousso:2012sj,Kelly:2013aja}) the natural sub-region to 
consider is the AdS causal wedge.
Consider a spherical region of an equal-time hypersurface on
$\Reals^{d-1,1}$:
\eq{
  \cR = 
  \left\{
    x \in \Reals^{d-1,1} \; \bigg| \; 
    x^0 = \overline{x}^0,\; \sum_{i=1}^{d-1} (x^i - \overline{x}^i)^2 = L 
  \right\} .
}
This region has origin $\overline{x}$ and radius $L$.
Let $D^{+(-)}_{\rm bndry}$ denote the future (past) domain of dependence on
$\Reals^{d-1,1}$. Then we refer the causal domain of dependence of $\cR$ as 
a causal development $\cD$:
\eq{
  \cD = D^+_{\rm bndry}[ \cR ] \cup D^-_{\rm bndry}[ \cR ] .
}
Suppose that $\cD$ is a sub-region on the
conformal boundary of the AdS Poincar\'e chart; then one may define
the associated AdS causal wedge
\eq{
  \cW = J^+_{\rm bulk}[ \cD ] \cap J^-_{\rm bulk}[ \cD ] ,
}
where $J^{+(-)}_{\rm bulk}$ is the future (past) domain of influence on
$d+1$-dimensional PAdS -- see Fig.~\ref{fig:wedge3d}. 
As with $\cR$ we say that the causal wedge $\cW$ has origin 
$\overline{x}$ and radius $L$.

The bulk region of any AdS causal wedge is isomorphic to the region
of AdS known as the AdS-Rindler wedge \cite{Casini:2011kv}.\footnote{The 
  AdS-Rindler wedge is also known 
  as AdS in hyperbolic coordinates and corresponds to the special case 
  of a hyperbolic AdS black hole metric when the geometry is in 
  fact just pure AdS \cite{Emparan:1999gf}.}
Thus the bulk of any causal wedge may be covered by the
AdS-Rindler coordinate chart which is independent of a particular wedge's
radius and origin. Of course, the radius and origin of a causal wedge is 
encoded in the diffeomorphism necessary to transform
from Poincar\'e coordinates to AdS-Rindler coordinates on the wedge.
This diffeomorphism induces a distinct conformal transformation on 
the boundary theory which similarly depends on the radius and origin of
the wedge. The details of this bulk diffeomorphism (boundary conformal 
transformation) is nicely 
described in \cite{Casini:2011kv}, and we do not need to repeat
them here. For completeness we provide an example of such of a 
transformation in Appendix~\ref{app:wedge}.
What is relevant for us is the fact that a boundary-to-bulk map
for the AdS-Rindler wedge immediately carries over to the more
general case of a causal wedge, and so we need only study the
former case in detail.

%%%%%%%%%%%%%%%%%%%%%%%%%%%%%%%%%%%%%%%%%%%%%%%%%%%%%%%%%%%%%%%%%
\subsection{The case of the AdS-Rindler wedge}
\label{sec:AdSR}
%%%%%%%%%%%%%%%%%%%%%%%%%%%%%%%%%%%%%%%%%%%%%%%%%%%%%%%%%%%%%%%%%

Without further ado we examine the boundary-to-bulk map for
the AdS-Rindler wedge.
To keep the notation light we will specialize 
our presentation to the case of $D = 2+1$ bulk dimensions. 
This also allows a more direct comparison with the recent
works \cite{Czech:2012be,Parikh:2012kg,Bousso:2012mh} as well
as the classic references \cite{Unruh:1984aa,Birrell:1982ix}.

In $2+1$ dimensions the AdS-Rindler metric has the line element
\eqn{  \label{eq:AdSRcoords}
  ds^2 &=& \frac{\ell^2}{z^2}\left[
    - \left(1-\frac{z^2}{\ell^2}\right) d\eta^2
    + \left(1-\frac{z^2}{\ell^2}\right)^{-1} dz^2
    + d\chi^2
  \right] .
}
It is useful to introduce a complete set of 
canonically-normalized bulk Klein-Gordon modes $S_k(X) = N_k e^{ik \cdot x}V_k(z)$ where
$k \cdot x = -\omega t + k^1 \chi$.
These modes are normalized such that two modes have Klein-Gordon 
inner product
\eqn{ \label{eq:KGIP}
  4\pi^2 \delta(\omega_1 -\omega_2)\delta(k^1_1 - k^1_2)
%  = -i \int d\Sigma n^\mu S_1(X) \overleftrightarrow{\nabla_\mu}
%  S_2^*(X) |_{\eta = {\rm const}}.
  = -i \int_{-\infty}^\infty d\chi \int_0^\ell \frac{dz}{z}
  \left(1-\frac{z^2}{\ell^2}\right)^{-1} S_1(X) \overleftrightarrow{\d_\eta} 
  S_2^*(X) \bigg|_{\eta = {\rm const}} .\quad
}
The $z$-dependence is contained in \cite{Hamilton:2006az}
\eqn{ \label{eq:AdSRV}
  V_k(z) 
  &=& 
  \left(\frac{z}{\ell}\right)^\Delta 
  \left(1 - \frac{z^2}{\ell^2} \right)^{-i\ell\omega/2}
  {}_2 F_1\hspace{-4pt}\left[\frac{\Delta-i\ell\omega +i\ell k^1}{2},
    \frac{\Delta-i\ell\omega-i\ell k^1}{2}
    ; \Delta ; \frac{z^2}{\ell^2}
  \right] .
}
The $V_k(z)$ are real, satisfy
\eq{
  V_{\omega,k^1}(z) = V_{-\omega,k^1}(z) = 
  V_{\omega,-k^1}(z) = V_{-\omega,-k^1}(z) ,
}
and are normalized such that near $z=0$ the modes behave as
\eq{
  V_k(z) = \left(\frac{z}{\ell}\right)^\Delta\left(1 + O(z^2)\right) ,
  \quad z \to 0 .
}
In order to satisfy (\ref{eq:KGIP}) the normalization constant 
must be chosen to satisfy\footnote{Recall
that the Gamma function satisfies $\Gamma(x)^* = \Gamma(x^*)$, so
it is easy to see that $|N_k|^2$ is indeed non-negative.}
\eqn{
  |N_k|^2 &=& \frac{1}{2 \ell |\omega|}
  \frac{
    \Gamma\left(\frac{\Delta + i\ell (\omega + k^1)}{2}\right)
    \Gamma\left(\frac{\Delta - i\ell (\omega + k^1)}{2}\right)
    \Gamma\left(\frac{\Delta + i\ell (\omega - k^1)}{2}\right)
    \Gamma\left(\frac{\Delta - i\ell (\omega - k^1)}{2}\right)
  }
  {\Gamma^2(\Delta)\Gamma(i\ell \omega)\Gamma(- i\ell \omega)} .
}

In this setting there are two natural vacua, each of which is
locally analytic on the wedge: the usual
global AdS vacuum $\Omega$ and the AdS-Rindler (AdSR) vacuum.
The latter is simply the state with zero particles in the 
particle basis defined by the AdS-Rindler coordinate $\eta$;
it has the Wightman 2-pt function
\eq{ \label{eq:AdSRW}
  \C{\Phi(X_1)\Phi(X_2)}_{\rm AdsR} =
  \ell \int \frac{d^2 k}{4\pi^2} \theta(\omega) |N_k|^2 
  e^{i k \cdot (x_1 - x_2)} V_k(z_1) V_k(z_2) .
}
This state induces on the boundary the Rindler (R) state 
whose 2-pt function is easily read off from (\ref{eq:AdSRW}):
\eq{ \label{eq:RonR}
  \C{\phi(x_1)\phi(x_2)}_{R} = 
  \ell^{2-2\Delta} \int \frac{d^2 k}{4\pi^2} 
  \theta(\omega) |N_k|^2 e^{i k \cdot (x_1 - x_2)} .
}
Returning to the bulk, consider now the global AdS vacuum 
restricted the the AdS-Rindler chart. As is well-known, 
on this chart the global AdS vacuum satisfies the KMS condition 
with inverse temperature $\beta = 2\pi\ell$ (in natural units):
\eq{
  \C{\Phi(\eta_1,x_1)\Phi(\eta_2,x_2)}_\Omega
  = \C{\Phi(\eta_2,x_2)\Phi(\eta_1+i 2\pi \ell,x_1)}_\Omega .
}
Using standard thermal field theory techniques
it is easy to deduce from this expression that the global AdS 
2-pt function may be written
% \footnote{For instance,
%   one may construct the Euclidean Green's function on the manifold
%   obtained by analytically continuing $\eta \to i\tau$ and letting 
%   $\tau$ have a period of $2\pi \ell$. Under Wick rotation this
%   Green's function becomes the time-ordered 2-pt function in the KMS
%   state.}
\eqn{ \label{eq:AdSRKMS}
  \C{\Phi(X_1)\Phi(X_2)}_\Omega
  &=&
  \ell \int \frac{d^2 k}{4\pi^2} e^{i k \cdot (x_1 - x_2)} 
  \frac{e^{\pi \ell \omega}}{2  |\sinh \pi \ell \omega|}
  |N_k|^2 V_k(z_1) V_k(z_2) .
}
This state induces the usual Minkowski vacuum state $\omega$
on the boundary:\footnote{This $\omega$ Wightman function may 
also be obtained by considering the action of the conformal 
transformation on the Minkowski 
vacuum 2-pt function in Cartesian coordinates. 
Under this conformal transformation the Minkowski 2-pt
function becomes
\eqn{
  \C{\phi(x_1)\phi(x_2)}_\omega
  &=& \frac{1}{2\pi}\frac{1}{(x_1 - x_1)^{2\Delta}}
  \to \frac{1}{(2\pi)2^\Delta \ell^{2\Delta}}
  \left(\cosh\left[\frac{\chi_1-\chi_2}{\ell}\right]
  -\cosh\left[\frac{\eta_1-\eta_2}{\ell}\right]\right)^{-\Delta}  .
    \label{eq:MonRx}
}
Taking the Fourier transform of this expression
yields (\ref{eq:MonRk}).
}
\eq{ \label{eq:MonRk}
  \C{\phi(x_1)\phi(x_2)}_{\omega} = 
  \ell^{2-2\Delta} \int \frac{d^2 k}{4\pi^2} 
  \frac{e^{\pi \ell \omega}}{2  |\sinh \pi \ell \omega|}
  |N_k|^2 e^{i k \cdot (x_1 - x_2)}  .
}
Like $\Omega$ in the bulk $\omega$ is a KMS state with 
$\beta = 2\pi\ell$. From the expressions above one may
easily verify the local analyticity of these states.

We can establish a map $\cAbndry(\cD) \mapsto \cAbulk(\cW)$
just as we did for the Poincar\'e chart.
The mapping kernel $K(X|y)$ is easily constructed from the bulk
Klein-Gordon modes:
\eq{
  K(X | y) = \frac{1}{\sqrt{-\gamma(y)}}
  \int \frac{d^2 k}{4\pi^2} e^{i k \cdot (x-y)} V_k(z) ,
}
where $\sqrt{\gamma(-y)} = \Omega^2(y) = e^{2\chi/\ell}$.
As we will see momentarily, the Klein-Gordon modes $V_k(z)$ can
grow as $|k| \to \infty$ and so $K(X|y)$ must be regarded as a
distribution. Formally $K(X|y)$ constructs a boundary test function 
$f_F(y)$ for every bulk test function $F(X) \in C^\infty_0(\cW)$ 
as in (\ref{eq:f_F}). However, recall from the discussion in
\S\ref{sec:boundary} that boundary test functions
yield well-defined observables only if they are members of the 
class $\cT$. Also recall that $f_F \in \cT$ only if $f_F(y)$ 
satisfies the wave front set condition (\ref{eq:cT}). Thus
our task is to determine if the $f_F(y)$ constructed using
$K(X|y)$ have wave front sets which satisfy (\ref{eq:cT}).

We can accomplish this by examining the behavior of the Fourier 
transforms $\hat{f}_F(k)$ in AdS-Rindler coordinates.
Of course the AdS-Rindler chart is not a locally-flat coordinate chart 
like the charts involved computing wave front sets, 
but nevertheless we may infer from the AdS-Rindler Fourier transform 
sufficient information to determine if $f_F \in \cT$.
To start consider a bulk test function
\eq{
  F(X) = \int \frac{d^2 k}{4\pi^2} e^{i k \cdot x} \hat{F}_k(z) ,
  \quad F \in C^\infty_0(\cW) .
}
Because $F(X)$ is smooth it follows that it's wave front set it
empty and we may convolve $F(X)$ with $K(X|y)$.
The resulting associated boundary test function $f_F(y)$ has the 
AdS-Rindler Fourier transform
\eq{
  \hat{f}_F(k) = 
  \ell^3 \int_0^\ell \frac{dz}{z^3} \hat{F}_k(z) V_k(z) .
}
We wish to determine the behavior of $\hat{f}_F(k)$ at large $|k|$,
and for this we need to know the behavior of $V_k(z)$ as
$|k| \to \infty$ in a given direction in momentum space while 
$\Delta$, $z$ are held fixed. Examining the 
equation of motion governing $V_k(z)$ one may readily see that this limit
corresponds to the limit where the effective mass-squared term becomes 
large in magnitude. 
In this regime the asymptotic form of $V_k(z)$ may be reliably
computed using the WKB approximation. We carry out this analysis
in Appendix~\ref{app:AdSRmodes}.
For $\omega^2 \gg (k^1)^2,\Delta^2$ we obtain the asymptotic form
\eq{ \label{eq:Vkapproxw}
  V_k(z) \approx \frac{2^{\Delta-1} \Gamma(\Delta)}
  {\sqrt{2\pi}} |\ell\omega|^{1/2-\Delta}
  \left(\frac{z}{\ell}\right)^{1/2}
  \left[\left(\frac{1+z/\ell}{1-z/\ell}\right)^{i \ell \omega/2}
    + \left(\frac{1+z/\ell}{1-z/\ell}\right)^{-i \ell \omega/2}\right] .
}
When $|k| \to \infty$ along a null direction there is 
qualitatively similar behavior, i.e. $V_k(z)$ is bounded by a 
power of $|k|$, with oscillatory dependence on $z/\ell$.
Finally, for $(k^1)^2 \gg \omega^2,\Delta$ we obtain the asymptotic form
\eq{ \label{eq:Vkapproxk}
  V_k(z) \approx C |\ell k^1|^{1-\Delta}
  \left[ \frac{z/\ell}{\ell|k^1|(1-(z/\ell)^2)^{1/2}}\right]^{1/2}
  \left( e^{\ell k^1\,{\rm arcsin}(z/\ell)} 
    + e^{-\ell k^1\, {\rm arcsin}(z/\ell)} \right) ,
}
where $C$ is a finite constant independent of $k^1$. This expression
diverges most strongly as $z \to \ell$, and from this behavior we
bound $V_k(z)$ in this regime by
\eq{
  V_k(z) \le C |k^1|^{1/2-\Delta} e^{\frac{\pi}{2}|k^1|} ,
}
with $C$ another finite constant independent of $k^1$.
Taken together, these results show that 
the AdS-Rindler Fourier transform $\hat{f}_F(y)$ decays faster than any 
power as $|k|\to\infty$ along timelike and null directions 
(in AdS-Rindler coordinates). 
This is sufficient to determine that the wave front
set of $f_F(y)$ satisfies the condition (\ref{eq:cT}), and
thus that $f_F \in \cT$.

Of course we do not need fancy wave front set arguments to show
that the boundary observables $\phi[f_F]$ have well-defined correlation
functions for physically reasonable states. One may obtain the same 
conclusion by direct computation, e.g. by examining the 
$\phi[f_F]$ correlators of $\omega$ and $R$.
Consider the correlator
\eq{
  \C{\phi(f_{F_1})\phi(f_{F_2})}_R 
  = \int \frac{d^2 k}{4\pi^2} |N_k|^2 \hat{f}_{F_1}(-k) \hat{f}_{F_2}(k) .
} 
Using the asymptotic formula for the Gamma function (\ref{eq:BigGamma}) 
we readily obtain the limits
\eqn{
  \label{eq:AdSR_Nklimit_k}
  \lim_{\ell |k^1| \to \infty} |N_k|^2
  &=& %\frac{2^{3-2\Delta}\pi}{\Gamma^2(\Delta)} |\sinh\pi\omega \ell|
  C_1 
  |\ell k^1|^{2\Delta-2} e^{- \pi \ell |k^1|} ,\quad
  \Delta, \; \omega \; \text{fixed},
  \\
  \label{eq:AdSR_Nklimit_w}
  \lim_{\ell |\omega| \to \infty} |N_k|^2
  &=& %\frac{2^{2-2\Delta}\pi}{\Gamma^2(\Delta)} 
  C_2
  |\ell \omega|^{2\Delta-2} ,
  \quad \Delta, \; k^1 \; \text{fixed},
}
with finite constants $C_1,C_2$. Combining this with the
the asymptotic behavior of the Klein-Gordon modes (\ref{eq:Vkapproxw}),
(\ref{eq:Vkapproxk}) we see that the Fourier transform of 
$\C{\phi(f_{F_1})\phi(f_{F_2})}_R$
converges absolutely. The story is quite the same for the 
Minkowski vacuum $\omega$ whose 2-pt function Fourier
transform differs only by the addition of a factor
$e^{\pi\ell\omega}/\sinh(\pi\ell\omega)$. The story is also the same
for states constructed from $\omega$ by acting with
$\cAbndry$ as well as states constructed from $R$ by acting with
$\cAbndry(\cD)$.

%%%%%%%%%%%%%%%%%%%%%%%%%%%%%%%%%%%%%%%%%%%%%%%%%%%%%%%%%%%%%%%%%
\section{Discussion}
\label{sec:disc}
%%%%%%%%%%%%%%%%%%%%%%%%%%%%%%%%%%%%%%%%%%%%%%%%%%%%%%%%%%%%%%%%%

The goal of this work has been to understand in a precise way
how ``bulk information'' is encoded in the boundary CFT using
the simple example of an AdS Klein-Gordon field as a case study.
In particular we have been interested in quantifying the
information available to a boundary observer with access limited 
to a sub-region of the boundary.
It is useful to describe our results in terms of three processes:
reading, writing, and storing information.
In \S\ref{sec:b2b} we showed that an observer with access to the
observables within domain of dependence $\cD$ on the boundary
may reconstruct all bulk observables contained within the associated
AdS causal wedge anchored to $\cD$. We say that this observer
may read the information of the bulk algebra of observables $\cAbulk(\cW)$.
Any observer with access to a boundary domain $\cD$ whose
associated causal wedge contains an observable $A \in \cAbulk$ may read
$A$. This set of boundary domains has no mutual overlap.
So, while the information of $A$ may be read from several boundary
domains, it is inappropriate to say that this information
is stored in any of them -- see again Fig.~\ref{fig:wedges}.

The ability of a boundary observer to write bulk information is 
encapsulated in the holographic form of the Reeh-Schlieder theorem:
\vspace{8pt}\\
{\it
Let $\cW$ be an AdS causal wedge with conformal boundary
$\cD$. Let $\Psi$ be locally analytic on $\cW$ and let $\psi$ be
it's boundary value. Then the linear spans of the sets of states
\eq{
  S_\cD = \left\{ A \psi \; | \; A \in \cAbndry(\cD) \right\} ,
  \quad
  S_\cW = \left\{ B \Psi \; | \; B \in \cAbulk(\cW) \right\} ,
}
satisfy ${\rm span}(S_\cW) \subset {\rm span}(S_\cD)$.}
\vspace{8pt}\\
By now the reader is familiar with the implications of this statement.
It we consider the case where $\Psi$ is the AdS vacuum $\Omega$ 
then the set $S_\cD$ spans the set of finite-particle states,
or more generally the set of finite energy-density states which
may be constructed from our bulk algebra of observables.
If we instead consider the vacuum state of a causal wedge 
$\cW_2 \supseteq \cW$
then set $S_\cD$ spans the full set of finite-particle states
built atop this vacuum. Either way, the point is that the
set of observables in $\cD$ is powerful enough to construct essentially
any state consistent with the bulk semi-classical approximation.
In this sense, by careful manipulation of $\cAbndry(\cD)$ one may
write more than one can read. 
Another interpretation of the bulk RS theorem is that 
%quantum field theory it is impossible
%to isolate a subsystem from outside influences.
%Said differently, 
states satisfying reasonable energy conditions
are have a high degree of entanglement. In holography we may 
exploit this entanglement to control the bulk system using only a 
sub-region of the boundary.
We remind the reader that we are considering here just the effects of
acting with normalizable excitations at the boundary.

At first glance our formulation of the Reeh-Schlieder theorem
appears somewhat weaker than that of Minkowski QFT found in 
textbooks \cite{Haag:1992aa,Streater:1989vi}. 
In textbook accounts the RS theorem states the the set of states
generated from the vacuum by a sub-set of the observable algebra
is dense on the Hilbert space.
Our theorem is phrased in terms of linear spans of sets of states
generated by a $*$-algebra.
The added power of the textbook theorems comes from additional
technical assumptions about the Hilbert space which follow from,
e.g., adopting the Wightman axioms as in the original work
\cite{Schlieder:1965aa} or some suitable generalization for curved
spacetimes \cite{Strohmaier:2002aa}.
We feel that our phrasing gets to the point of
the RS theorem without boring the reader with too many technical details.
That said, we are happy to point out a few technical assumptions
of our analysis which could be violated so as to nullify our conclusions:
\begin{enumerate}
  \item We consider $*$-algebras of observables which have no norm.
    Using these algebras we cannot 
    generate an infinite particle state from a state with finitely many
    particles, nor can we construct compactly-supported unitary operators.
    The latter are especially useful for avoiding the conclusions of 
    the RS theorem. In order to describe these objects 
    consistently one has do to more work and specify a norm for 
    the algebras (equivalently, a norm for the Hilbert space).
  \item We restrict attention to states which may be generated
    from locally analytic states. We are motivated to do so because
    we know that these states have well-behaved average stress-energy
    fluctuations. Of course one could consider other, more singular states. 
\end{enumerate}

\begin{figure}
  \centering
  \ifcolorpics
  \includegraphics[width=0.5\textwidth]{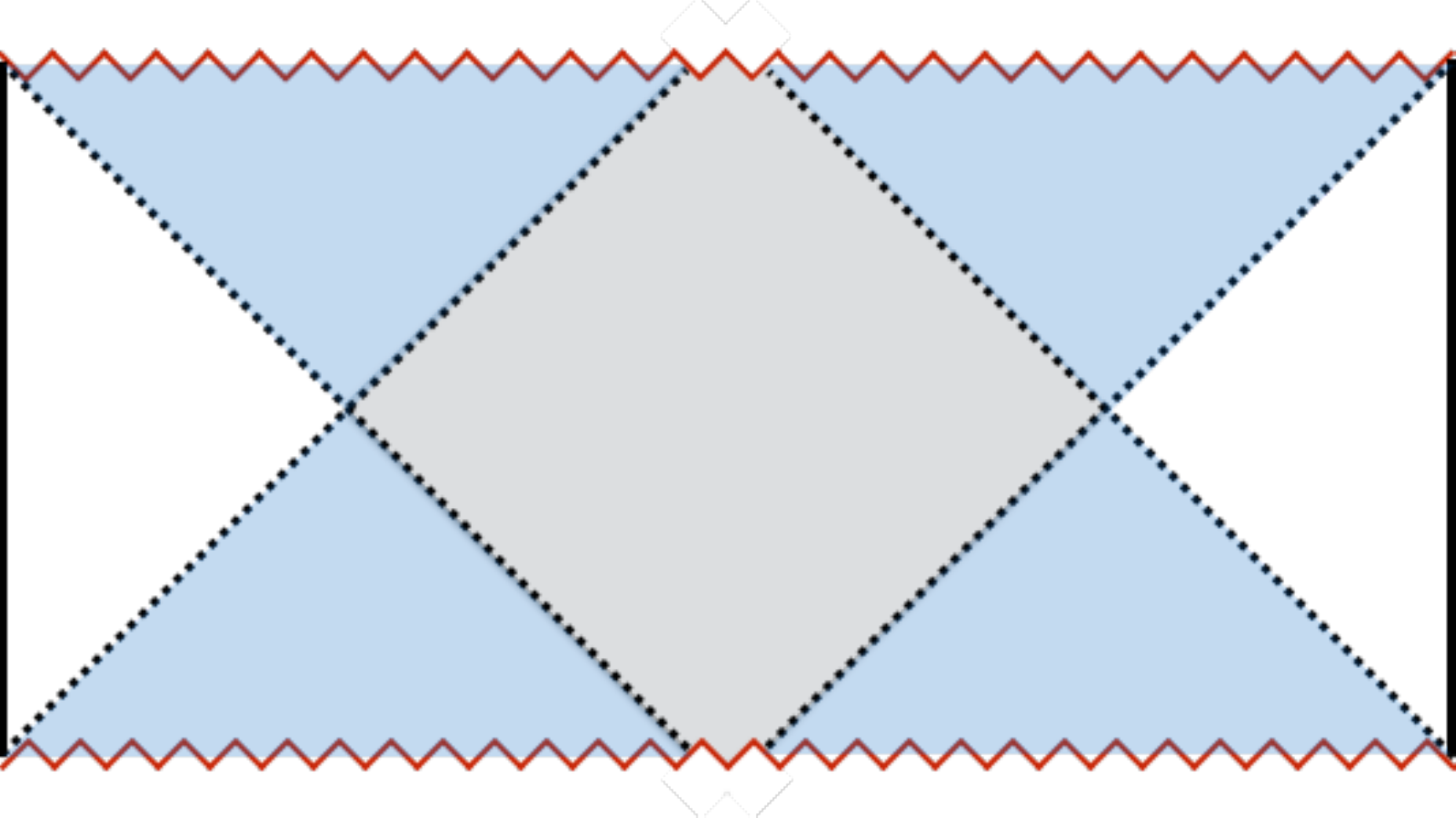}
  \else
  \includegraphics[width=0.5\textwidth]{AdSwormholebw}
  \fi
  \caption{Conformal diagram of a simple AdS wormhole geometry.
    There are two disconnected conformal boundaries.
    Each blue shaded region has causal contact with the boundary
    only through one of it's past or future domains of influence.
    The gray shaded region has no causal contact with the boundaries.
    \label{fig:wormhole}}
\end{figure}

It is natural to ask how our analysis might be extended to more 
interesting bulk geometries. To make this discussion as interesting as
possible let us consider asymptotically-AdS wormhole geometries
\cite{Aminneborg:1997aa,Krasnov:2000zq,Krasnov:2003aa,Skenderis:2009ju,
  Shenker:2013yza},
an example of which is depicted in Fig.~\ref{fig:wormhole}.
These geometries have regions of the bulk with no causal contact
with the boundaries.
We continue to operate within the semi-classical approximation
where we have a known, fixed bulk geometry and would like to
recover the bulk observables of a bulk scalar field given only
it's boundary correlation functions.
First consider recovery for bulk observables in the regions exterior
to the wormhole (the white regions in Fig.~\ref{fig:wormhole}).
Given our analysis of bulk-to-boundary maps in exact AdS, along with
the many excellent previous works on this subject
\cite{Bena:1999jv,Hamilton:2005ju,Hamilton:2006az,Hamilton:2006fh,
  Lowe:2008ra,Kabat:2011rz,Heemskerk:2012mq,Kabat:2012hp,Kabat:2012av},
we see no credible obstacle to constructing such a map between
a suitable algebra of boundary observables and the algebra of 
bulk observables restricted to these exterior regions.
But such a map does not tell us about the quantum state beyond 
the horizon; for this we need additional ingredients.

Recently some works \cite{Papadodimas:2012aq,Verlinde:2013qya}
which consider asymptotically-AdS black holes
have advocated analytic continuation as a possible tool for 
extending a state defined in the exterior
region to the region past the horizon.
Our concern with this approach is that such an analytic continuation
does not necessarily provide the unique extension of a state.
As we mentioned when discussing locally analytic states in AdS,
it is very rare to have a unique notion of analytic continuation for
quantum states.
Even in AdS very few states could be 
called analytic from any reasonable perspective. Leaving aside 
their distributional aspects, the correlation functions
of typical well-behaved quantum states are at best smooth,
and there exist many ways to extend a smooth solution to a local
equation of motion from a given domain into it's causal complement.
In this respect we sympathize with \cite{Avery:2013bea}:
proving the existence of a smooth extension of correlation functions
past the horizon is not the same thing a showing that
the physical state in question corresponds to this extension.

Perhaps a more promising approach is to take advantage of the
additional information available to us if we have access to a complete
Cauchy surface of the boundary theory. In this case we have access 
to any global charge of the bulk theory which satisfies a Gauss' Law 
type conservation law. 
As has been emphasized by many before
\cite{Balasubramanian:2006aa,Balasubramanian:2006ab,Marolf:2008mf,Marolf:2008mg,Marolf:2013iba},
theories with dynamical gravity and stringy excitations necessarily
have such charges, including the bulk gravitational Hamiltonian.
Within the semi-classical setting one can imagine toy 
quantum field theories on AdS wormhole geometries
for which there exist enough Gauss' law-type charges to completely 
determine the bulk quantum state. It would be very interesting
to study such models in detail.\\

\noindent{\bf Acknowledgments} \\

We thank Madeline Anthonisen, Robert Brandenberger, Daniel Kabat, Alex Maloney, 
Guy Moore, Matthew Roberts, Vladimir Rosenhaus, and Aron Wall for 
useful conversations.

% \cite{Polchinski:1999aa,Suskind:2000ab}% precursors
% \cite{Marolf:2012xe}% superselection sectors
% \cite{Maldacena:2001kr}% bh paper
% \cite{Kay:2013gia}

%%%%%%%%%%%%%%%%%%%%%%%%%%%%%%%%%%%%%%%%%%%%%%%%%%%%%%%%%%%%%%%%%
\appendix
%%%%%%%%%%%%%%%%%%%%%%%%%%%%%%%%%%%%%%%%%%%%%%%%%%%%%%%%%%%%%%%%%

%%%%%%%%%%%%%%%%%%%%%%%%%%%%%%%%%%%%%%%%%%%%%%%%%%%%%%%%%%%%%%%%%
\section{Wave front sets}
\label{app:WF}
%%%%%%%%%%%%%%%%%%%%%%%%%%%%%%%%%%%%%%%%%%%%%%%%%%%%%%%%%%%%%%%%%

In this appendix we provide a brief introduction to the 
wave front set of a distribution, following closely the standard
reference \cite{Hormander:1990aa} as well as the introduction
contained in \cite{Brunetti:1995rf}.
We denote the set of smooth functions on a manifold $M$ by
$\mathcal{E}(M)$ and the corresponding dual space of distributions
$\mathcal{E}'(M)$. 
We are interested in characterizing the singularity structure of 
a distribution $u \in \mathcal{E}'(M)$. 
For the moment we consider $M = \Reals^n$.
We denote the Fourier transform of $u$ by $\hat{u}$, and we refer 
to the dual variables of the Fourier transform as position 
$x$ and momentum $k$ respectively.

The most basic measure of singularity in position space is the 
\emph{singular support} of $u$, denoted ${\rm singsupp}(u)$, which is the
set of points in $\Reals^n$ having no open neighborhood to 
which the restriction of $u$ is a smooth function. 
The singular support describes where in position space a distribution
is singular, but in no way describes ``how'' it is singular.

We may also describe the singular nature of a distribution
in momentum space. Recall that the Fourier 
transform of a smooth function $f \in \mathcal{E}(\Reals^n)$ decays 
faster than any power law at large momenta:
\eq{ \label{eq:smooth}
  |\hat{f}(k)| \le \frac{C_N}{(1+|k|)^N} , \quad N \in \mathbb{N}_0 ,
}
for some finite constants $C_N$.
Unless $u$ is in fact induced by a smooth function $\hat{u}$ will
fail the bound (\ref{eq:smooth}) in some directions of momentum space.
We may thus define the \emph{singular cone} of $u$, denoted $\Sigma(u)$,
as the conic set\footnote{
  A conic set $V \subset \Reals^n \setminus \{0\}$ 
  is a set of points containing $k$ as well as all points
  $ c k$ for all constants $c > 0$.
} of all $k \in \Reals^n \setminus \{ 0 \}$ having no conic 
neighborhood for which (\ref{eq:smooth}) is valid.
The singular cone describes the ``singular directions'' in momentum 
space for which $\hat{u}$ fails to behave like the Fourier
transform of a smooth function.

N every direction $k \in \Sigma(u)$ is effectively
problematic at a given point $x \in {\rm singsupp}(u)$. 
In order to determine which singular directions are the culprits of
a given singularity we need a more refined, local notion of the 
singular directions at a given point. 
A basic property of the singular cone is that 
\eq{
  \Sigma(f u) \subseteq \Sigma(u) , \quad f \in \mathcal{E}(\Reals^n) .
}
This gives us a natural way to define the cone of singular directions
at a point $x$:
\eq{
  \Sigma_x(u) := \cap_f \Sigma(f u), \quad
  f \in \mathcal{E}(\Reals^n) ,\quad f(x) \neq  0 .
}
We call the pair $(x_0,k_0) \in \Reals^n \times (\Reals^n \setminus \{0\})$
a \emph{singular directed point} of $u$ if $x_0 \in {\rm singsupp}(u)$
and $k_0 \in \Sigma_{x_0}(u)$. The set of singular directed points of
$u$ is it's wave front set:

\begin{definition}
  The wave front set of a distribution 
  $u \in \mathcal{E}'(\Reals^n)$ is
  \eq{
    \WF(u) := \{ (x,k) \in \Reals^n \times (\Reals^n \setminus \{0\}) \; | \;
    k \in \Sigma_x(u) \} .
  }
\end{definition}

\noindent The wave front set is a conic set with respect to the momentum
variable. The projection of $\WF(u)$ in position space is ${\rm singsupp}(u)$;
the projection of $\WF(u)$ in momentum space is $\Sigma(u)$. 
The wave front set cannot be enlarged by convolution with a smooth 
function,
\eq{
  \WF(f u) \subseteq \WF(u), \quad f \in \mathcal{E}(\Reals^n), 
}
or by a derivative operator $\d$:
\eq{
  \WF(\d u) \subset \WF(u) .
}\\

\begin{example}
The Dirac delta function $\delta(x)$ on $\Reals^n$ has the wave front set
\eq{
  \WF(\delta) = \left\{ (x,k) \in \Reals^n \times \Reals^n
    \setminus \{ 0 \} \; \big| \; x = 0 
  \right\} .
}
\end{example}

The wave front set provides a precise criteria for when the point-wise
product of a pair of distributions is well-defined (cf.
Theorem 8.2.10 of \cite{Hormander:1990aa}). The point-wise product
$u v$ of two distributions $u$, $v$ is uniquely defined 
unless there exists a singular directed point $(x,k)$ such 
that $(x,k) \in \WF(u)$ and $(x,-k) \in \WF(v)$. 
Roughly speaking, this criteria states that the convolution of
distributions unambiguously defines a new distribution so
long as singular directed points of do not overlap in convolution.

We have been discussing the $C^\infty$ wave front set
which describes the failure of a distribution to be smooth. One
can similarly define an \emph{analytic wave front set} ($\WF_A$)
which describes the failure of a distribution to be analytic.
For details see \cite{Hormander:1990aa}. Obviously, 
$\WF(u) \subset \WFA(u)$.

The wave front set generalizes quite easily to distributions on manifolds.
To see this note that both the singular
support and the singular cone at $x$ are locally
defined, and thus the wave front set is a local concept. It can
be shown that the wave front set transforms covariantly under
diffeomorphisms. Taken together these properties indicate that
the wave front set is properly thought of as a conic subset of the
cotangent bundle, i.e. $\WF \subset T^*M \setminus \{ 0 \}$.
The properties of the wave front set described above for 
above hold as well for 
$u \in \mathcal{E}'(M)$, $f \in \mathcal{E}(M)$, 
where $M$ is an arbitrary manifold.
Further details of wave front sets for distributions on manifolds 
may be found in, e.g., 
\cite{Radzikowski:1996aa,Brunetti:1995rf,Verch:1999aa}.

\begin{example} 
  Consider a CFT on $\Reals^{d,1}$.
  The 2-pt function of a scalar CFT operator of weight $\Delta$ with
  respect to the CFT vacuum is given by
  \eq{
    W(x_1,x_2) \propto (x_1 -x_2)^{-2\Delta} ,
  }
  with the $i\epsilon$ prescription $x_2^0 \to x_2^0 -i\epsilon$.
  The wave front set of this 2-pt function is
  \eq{
    \WF(W) = \left\{ (x_1,k_1;x_2,-k_1) \in (T^*(\Reals^{d,1}))^2
      \setminus \{ 0 \} \; \Big| \; (x_1 - x_2)^2 = 0, \; 
      k_1 \in V^+    \right\} .
  }
  The closed forward lightcone $V^+$ is defined below equation
  (\ref{eq:WFOmega}).
\end{example}

%%%%%%%%%%%%%%%%%%%%%%%%%%%%%%%%%%%%%%%%%%%%%%%%%%%%%%%%%%%%%%%%%
\section{An example transformation to an AdS causal wedge}
\label{app:wedge}
%%%%%%%%%%%%%%%%%%%%%%%%%%%%%%%%%%%%%%%%%%%%%%%%%%%%%%%%%%%%%%%%%

In this Appendix we give an example of the bulk diffeomorphism
(boundary conformal transformation) which establishes an AdS-Rindler
chart on an AdS causal wedge. Although the procedure is the same
in all bulk dimensions $D = d+1 \ge 3$ the notation is least 
cumbersome if we restrict to $D = 2+1$ bulk dimensions.

Recall that AdS${}_3$ may be defined as the single-sheet hyperbaloid
of radius $\ell$ in a $\Reals^{2,2}$ embedding space
\eq{ \label{eq:AdSembedding}
  \text{AdS}_{3} := \left\{ \cX \in \Reals^{2,2} \; | \;
    -(\cX^{-1})^2 - (\cX^{0})^2 + (\cX^{1})^2 + (\cX^{2})^2 
    = -\ell^2  \right\} .
}
The embedding coordinates $\cX$ provide a very useful way of relating
AdS coordinate charts. The Poincar\'e chart
(\ref{eq:AdSPoincare}) may be related to the embedding coordinates
via
\eqn{ \label{eq:AdSPembedding}
  \cX^{-1} &=& \ell \frac{T}{Z} , \nn \\
  \cX^{0} &=& \frac{1}{2 Z} \left(-T^2 +Y^2 +Z^2 +\ell^2\right) ,\nn \\
  \cX^{1} &=& \frac{1}{2 Z} \left( T^2 -Y^2 -Z^2 +\ell^2\right) ,\nn \\
  \cX^{2} &=& \ell \frac{Y}{Z} ,
}
and the AdS-Rindler chart (\ref{eq:AdSRcoords}) is related to the embedding
coordinates via
\eqn{ \label{eq:AdSRembedding}
  \cX^{-1} &=& \frac{\ell^2}{z} \left(1-\frac{z^2}{\ell^2}\right)^{1/2}
  \sinh \frac{\eta}{\ell} , \nn \\
  \cX^{0} &=& \frac{\ell^2}{z} \cosh\frac{\chi}{\ell} , \nn \\
  \cX^{1} &=& - \frac{\ell^2}{z} \sinh\frac{\chi}{\ell} , \nn \\
  \cX^{2} &=& \frac{\ell^2}{z} \left(1-\frac{z^2}{\ell^2}\right)^{1/2}
  \cosh \frac{\eta}{\ell} .
}
In the embedding space description the AdS isometries are the subset
of isometries of $\Reals^{2,2}$ which preserve the hyperbaloid, i.e.
they are the boosts and rotations of $\Reals^{2,2}$. Consider a boost
in the $\cX^{0}$-$\cX^2$ plane. Although this is an AdS isometry it
does not preserve the AdS-Rindler chart (\ref{eq:AdSRembedding}).
Thus, for each boost parameter $\beta$ we may define boosted 
embedding coordinates $\cX(\beta)$ and then a new AdS-Rindler chart 
defined as in (\ref{eq:AdSRembedding}) but with $\cX$ on the left-hand side
replaced by $\cX(\beta)$. Relating this new AdS-Rindler chart to the
original embedding coordinates $\cX$ we obtain
\eqn{ \label{eq:AdSRBembedding}
  \cX^{-1} &=& \frac{\ell^2}{z} \left(1-\frac{z^2}{\ell^2}\right)^{1/2}
  \sinh \frac{\eta}{\ell} , \nn \\
  \cX^{0} &=&  \frac{\ell^2}{z} \left[\cosh\beta \cosh\frac{\chi}{\ell} 
  + \left(1-\frac{z^2}{\ell^2}\right)^{1/2}
  \sinh\beta \cosh \frac{\eta}{\ell} \right]
  , \nn \\
  \cX^{1} &=& - \frac{\ell^2}{z} \sinh\frac{\chi}{\ell} , \nn \\
  \cX^{2} &=& \frac{\ell^2}{z} \left[\sinh\beta \cosh\frac{\chi}{\ell}
  + 
   \left(1-\frac{z^2}{\ell^2}\right)^{1/2}
   \cosh\beta \cosh \frac{\eta}{\ell} \right] .
}
One may now use (\ref{eq:AdSPembedding}) to relate the coordinates 
defined on the right-hand side of (\ref{eq:AdSRBembedding}) to the
original Poincar\'e chart. The resulting expressions for $(T,Y,Z)$ in
terms of $(\eta,\chi,z)$ are complicated; what is important is to note
that near the boundary $z \to 0$ these relations are
\eqn{
  T &=& \frac{ \ell \sinh \frac{\eta}{\ell}}
  {\cosh\beta \cosh \frac{\chi}{\ell} + \sinh\beta \cosh\frac{\eta}{\ell}
  - \sinh\frac{\chi}{\ell}} + O(z^2), \nn \\
  Y &=& \frac{ \ell \left(
      \cosh\beta\cosh\frac{\eta}{\ell} + \sinh\beta\cosh\frac{\chi}{\ell}
    \right)}
  {\cosh\beta \cosh \frac{\chi}{\ell} + \sinh\beta \cosh\frac{\eta}{\ell}
  - \sinh\frac{\chi}{\ell}} + O(z^2) .
}
Thus, after converting bulk coordinates from Poincar\'e to the boosted
AdS-Rindler system the boundary metric takes the form
\eq{
  ds^2_{\rm bndry} = \Omega^2(x) \left( -d\eta^2 + d\chi^2 \right),
}
where the conformal factor is
\eq{ \label{eq:OmegaWedge}
  \Omega(x) = \frac{1}{\cosh\beta\cosh\frac{\chi}{\ell}
    +\sinh\beta\cosh\frac{\eta}{\ell}-\sinh\frac{\chi}{\ell}
  } .
}
This new boundary chart covers a domain of dependence $\cD$
with origin $\overline{x} = (0,\ell\coth\beta)$ and radius 
$L =\ell\,{\rm csch}\,\beta$.
In the bulk the horizon at $z = \ell$ corresponds precisely to the
edge of the associated spherical causal wedge.
In a similar fashion one may use a second boost in the embedding space
to shift the location of the origin of the causal wedge.

%%%%%%%%%%%%%%%%%%%%%%%%%%%%%%%%%%%%%%%%%%%%%%%%%%%%%%%%%%%%%%%%%
\section{WKB approximations of AdS-Rindler Klein-Gordon modes}
\label{app:AdSRmodes}
%%%%%%%%%%%%%%%%%%%%%%%%%%%%%%%%%%%%%%%%%%%%%%%%%%%%%%%%%%%%%%%%%

Here we derive the WKB approximations for the AdS-Rindler Klein-Gordon
mode functions quoted in \S\ref{sec:AdSR}. In AdS-Rindler coordinates
(\ref{eq:AdSRcoords}) the scalar d'Alembertian is
\eq{
  \Box \Phi(X) = \frac{z^2}{\ell^2} \left[
    -\left(1-\frac{z^2}{\ell^2}\right)^{-1}
    \d_\eta^2 + \d_\chi^2 + \left(1-\frac{z^2}{\ell^2}\right) \d_z^2
    - \left(1-\frac{z^2}{\ell^2}\right)\frac{1}{z}\d_z
    \right] \Phi(X) .
}
Using this and the form of the Klein-Gordon modes
$e^{ik \cdot x}V_k(z)$ we obtain the equation satisfied by the radial
function $V_k(z)$. To keep the notation manageable we adopt 
the dimensionless variables $\hat{z} = z/\ell$, $\hat{k}^1 = 
\ell k^1$, and $\hat{\omega} = \ell \omega$. Then the equation
for $V_k(z)$ may be written
\eq{ \label{eq:VEOM}
  (1-\hat{z}^2) V_k''(\hat{z}) - \frac{1+\hat{z}^2}{\hat{z}} V_k'(\hat{z})
  -  \cM^2(\hat{z}) V_k(\hat{z}) = 0 ,
}
where we have defined the effective mass-squared
\eq{ 
  \cM^2(\hat{z}) = \frac{\Delta(\Delta-2)}{\hat{z}^2} + (\hat{k}^1)^2 
    - \frac{\hat{\omega}^2}{1-\hat{z}^2}  .
}
We are interested in three regimes:
\begin{itemize}
  \item {\bf case i):} the large frequency regime
    \eq{
      \hat{\omega}^2 \gg (1-\hat{z}^2) 
      \left[ \frac{\D(\D-2)}{\hat{z}^2} + (\hat{k}^1)^2\right] ,
    }
  \item {\bf case ii):} the large spatial momentum regime
    \eq{
      (\hat{k}^1)^2 \gg 
      \frac{\hat{\omega}^2}{1-\hat{z}^2} - \frac{\Delta(\Delta-2)}{\hat{z}^2} ,
    }
    and
  \item {\bf case iii):} large null momentum, e.g., for
    $\hat{k}^\pm = \hat{\omega} \pm \hat{k}^1$
    \eq{
      \frac{(\hat{k}^+)^2}{4}  \gg - \frac{(\hat{k}^-)^2}{4} +
      \frac{(1-\hat{z}^2)}{\hat{z}^2}\left[
        \frac{\Delta(\Delta-2)}{\hat{z}^2} + O(\hat{k}^\pm)
      \right] .
    }
\end{itemize}
For each of these regimes the effective mass-squared $|\cM^2(\hat{z})| \gg 1$,
so we may use a WKB analysis to obtain an approximate solution.
We anticipate a solution of the form
\eq{ \label{eq:WKB1}
  V(\hat{z}) = K e^{A(\hat{z})}, 
}
where $A(\hat{z}) = A_0(\hat{z}) + A_1(\hat{z}) + \dots$ contains successive
terms suppressed by a large factor $\sim \cM(\hat{z})$. 
Under this assumption we may insert (\ref{eq:WKB1}) into (\ref{eq:VEOM}) 
and solve order-by-order for the $A_i(\hat{z})$. At lowest order only
the term $(A_0'(\hat{z}))^2$ contributes and we obtain
\eq{
  A_0'(\hat{z}) = \pm \frac{\cM(\hat{z})}{(1-\hat{z}^2)^{1/2}} ,
}
where $\cM(\hat{z})$ may be real or imaginary depending on the regime
of interest. At the next order we determine that
\eq{
  A_1(\hat{z}) = \half \ln \left[\pm \frac{\hat{z}}{(1-\hat{z}^2)^{1/2}\cM(\hat{z})} \right].
}
To proceed further we specialize for each case: \\

\noindent {\bf Case i):} Here the frequency term dominates the effective
mass-squared, making the mass imaginary:
\eq{
  \cM(\hat{z}) \approx \pm i \frac{\omega}{\sqrt{1-\hat{z}^2}} .
}
Inserting this into the results above, and integrating to obtain
$A_0(\hat{z})$, we have
\eq{ 
  A_0(\hat{z}) = \pm i \frac{\hat{\omega}}{2} \ln\left[\frac{1+\hat{z}}{1-\hat{z}}\right],
  \quad
  A_1(\hat{z}) = \half \ln\left[ \pm i \frac{\hat{z}}{\hat{\omega}}\right] .
}
Thus our approximate solution in this regime is
\eq{ \label{eq:AdSRWKB1}
  V_k(\hat{z}) \approx \left(\frac{\hat{z}}{|\hat{\omega}|}\right)^{1/2}
  \left[K_k \left(\frac{1+\hat{z}}{1-\hat{z}}\right)^{i \hat{\omega}/2}
    + K_k^* \left(\frac{1+\hat{z}}{1-\hat{z}}\right)^{-i \hat{\omega}/2}\right] .
}
We have combined solutions so that $V_k(\hat{z}) \in \Reals$.

The accuracy of our WKB approximation is governed by the ratio
$| A_1(\hat{z}) / A_0(\hat{z})|$ (as well as similar expressions at higher
orders). From the results above we see that this ratio is suppressed by 
a large factor $|\ln\hat{\omega} / \hat{\omega}|$ as expected, and
thus the WKB approximation is valid as $\hat{\omega}^2 \to \infty$
for fixed $\Delta$, $\hat{k}^1$, and $\hat{z}$.
However, $| A_1(\hat{z}) / A_0(\hat{z})|$ is also $\hat{z}$-dependent.
For parametrically small $\hat{z} \sim \hat{\omega}^{-1}$ this ratio 
becomes of order unity and the WKB approximation breaks down.
On the bright side, for $\hat{z}$ above this threshold the WKB approximation
is valid and even becomes exact as $\hat{z} \to 1$ 
($| A_1(\hat{z}) / A_0(\hat{z})| \to 0$ as $\hat{z} \to 1$).
Therefore the approximate solution (\ref{eq:AdSRWKB1}) can be matched 
to the exact solution in this limit, and in this way we can determine
the coefficient $K_k$.
For $\hat{z} = 1-\epsilon$ with $\epsilon \ll 1$ the exact result
is
\eqn{ \label{eq:match1}
  V_k(\hat{z} = 1-\epsilon) &=& \left[
  \frac{\Gamma(\Delta)\Gamma(i\hat{\omega})}
  {\Gamma\left(\frac{\Delta+i\hat{\omega}+i\hat{k}^1}{2}\right)
  \Gamma\left(\frac{\Delta+i\hat{\omega}-i\hat{k}^1}{2}\right)} 
  \epsilon^{-i\hat{\omega}/2}
  + (\hat{\omega} \to -\hat{\omega})
  \right]
  (1 + O(\epsilon)) .
}
Using this expression and the asymptotic form of the Gamma function
\eq{ \label{eq:BigGamma}
  \lim_{|y| \to \infty} \Gamma(x+iy)
  = (2\pi)^{1/2} e^{-(\pi/2)|y|} |y|^{x-1/2}\left(1 + O(y^{-1})\right) ,
}
we determine that the coefficient $K_k$ is
\eq{
  K_k = K_k^* = \frac{2^{\Delta-1} \Gamma(\Delta)}
  {\sqrt{2\pi}} |\hat{\omega}|^{1-\Delta} ,
}
as quoted in the text.

Note that one could also determine the $\hat{\omega}$ dependence
of the coefficient by matching the WKB approximation to the exact
solution in the region $\hat{z} \sim \hat{\omega}^{-1}$. The lack of
precision in this matching is reflected by the fact that we can only
determine the constant up to an $O(1)$ coefficient independent of 
$\hat{\omega}$. Through this approach one obtains 
$K_k  = C |\hat{\omega}|^{1-\Delta}$, which is of course consistent
with the (more accurate) result found above. \\

\noindent{\bf Case ii):} Now we specialize to case the large momentum
regime for which
\eq{
  \cM(\hat{z}) \approx \pm \hat{k}^1 .
}
Then
\eq{
  A_0(\hat{z}) = \pm \hat{k}^1 {\rm arcsin\,} \hat{z} ,
  \quad
  A_1(\hat{z}) = \half \ln \left[ \pm \frac{\hat{z}}{(1-\hat{z}^2)^{1/2} \hat{k}^1} \right], 
}
and our approximate solution becomes
\eq{ \label{eq:AdSRWKB2}
  V_k(\hat{z}) \approx K_k
  \left[ \frac{\hat{z}}{|\hat{k}^1|(1-\hat{z}^2)^{1/2}}\right]^{1/2}
  \left( e^{\hat{k}^1\,{\rm arcsin\,}\hat{z}} + e^{-\hat{k}^1\, {\rm arcsin\,}\hat{z}} \right) .
}
The ratio $|A_1(\hat{z})/A_0(\hat{z})|$ that controls the accuracy
of this approximation is suppressed by a factor of
$|\ln \hat{k}^1 / \hat{k}^1|$ and so the approximation is good for fixed
$\Delta$, $\hat{z}$. The $\hat{z}$-dependence of the ratio
causes it to become order unity when $\hat{z} \sim (\hat{k}^1)^{-1}$.
We may therefore match (\ref{eq:AdSRWKB2}) to the exact $V_k(z)$ 
in this region and determine that
\eq{
  K_k = C |\hat{k}^1|^{1-\Delta} ,
}
for constant $C$ independent of $\hat{\omega}$ and $\hat{k}^1$.
We note that the WKB approximation formally breaks down when
$1-\hat{z} \sim \exp[- c |\hat{k}^1|]$ as the ratio
$|A_1(\hat{z})/A_0(\hat{z})|$ once again becomes order unity,
but despite this as far as the dependence on $\hat{k}^1$ is
concerned there is excellent agreement between (\ref{eq:AdSRWKB2}) 
and (\ref{eq:match1}). \\

\noindent{\bf Case iii):}
Finally we describe when $|k| \to \infty$ is taken
along a null direction. This case is qualitatively similar to
the large frequency regime (i). Consider when
$|\hat{k}^+| \gg |\hat{k}^-|,\Delta$,$\hat{z}$; then the effective mass is
\eq{
  \cM(\hat{z}) \approx \pm \frac{i}{2} 
  \frac{\hat{z} \hat{k}^+}{\sqrt{1-\hat{z}^2}} ,
}
and so
\eq{
  A_0(\hat{z}) = \pm i \frac{\hat{k}^+}{4} \ln(1-\hat{z}^2),
  \quad
  A_1(\hat{z}) = - \half \ln\left[\mp i \frac{\hat{k}^+}{2} \right] ,
}
and our WKB approximation is
\eq{ \label{eq:AdSRWKB3}
  V(\hat{z}) \approx \left|\hat{k}^+\right|^{-1/2}
  \left[ K_k \left(1-\hat{z}^2\right)^{+ i k^+ / 4} +
    K_k^* \left(1-\hat{z}^2\right)^{- i k^+/4 }
  \right] .
}
The ratio $|A_1(\hat{z})/A_0(\hat{z})| \to 0$ as $\hat{z} \to 1$
indicating exact agreement in this limit, so we match (\ref{eq:AdSRWKB3})
with the exact expression (\ref{eq:match1}) to determine the coefficient
\eq{
  K_k = \frac{2^{\Delta/2}\Gamma(\Delta)}
  {\Gamma\left(\frac{\Delta+i \hat{k}^-}{2}\right)}
  |\hat{k}^+|^{(1-\Delta)/2} .  
}

%%%%%%%%%%%%%%%%%%%%%%%%%%%%%%%%%%%%%%%%%%%%%%%%%%%%%%%%%%%%%%%%%
%% Bibliography
%%%%%%%%%%%%%%%%%%%%%%%%%%%%%%%%%%%%%%%%%%%%%%%%%%%%%%%%%%%%%%%%%

\addcontentsline{toc}{section}{References}
\bibliographystyle{utphys}
\bibliography{AdSwedgeBib}

\providecommand{\href}[2]{#2}\begingroup\raggedright\begin{thebibliography}{10}

\bibitem{Schlieder:1965aa}
S.~Schlieder, ``{Some Remarks about the Localization of States in a Quantum
  Field Theory},'' \href{http://dx.doi.org/10.1007/BF01645904}{{\em Comm. Math.
  Phys.} {\bfseries 1} (1965) 265--280}.

\bibitem{Haag:1992aa}
R.~Haag, {\em Local quantum physics: fields, particles, algebras}.
\newblock Texts and monographs in physics. Springer-Verlag, 1992.
\newblock 390 p.

\bibitem{Aharony:1999ti}
O.~Aharony, S.~S. Gubser, J.~M. Maldacena, H.~Ooguri, and Y.~Oz, ``{Large N
  field theories, string theory and gravity},''
  \href{http://dx.doi.org/10.1016/S0370-1573(99)00083-6}{{\em Phys. Rept.}
  {\bfseries 323} (2000) 183--386},
\href{http://arxiv.org/abs/hep-th/9905111}{{\ttfamily arXiv:hep-th/9905111}}.
%%CITATION = HEP-TH/9905111;%%.

\bibitem{DHoker:2002aw}
E.~D'Hoker and D.~Z. Freedman, ``{Supersymmetric gauge theories and the AdS/CFT
  correspondence},''
\href{http://arxiv.org/abs/hep-th/0201253}{{\ttfamily ArXiv:hep-th/0201253}}.
%%CITATION = HEP-TH/0201253;%%.

\bibitem{Horowitz:2006ct}
G.~T. Horowitz and J.~Polchinski, ``{Gauge/gravity duality},''
\href{http://arxiv.org/abs/gr-qc/0602037}{{\ttfamily arXiv:gr-qc/0602037
  [gr-qc]}}.
%%CITATION = GR-QC/0602037;%%.

\bibitem{Czech:2012bh}
B.~Czech, J.~L. Karczmarek, F.~Nogueira, and M.~Van~Raamsdonk, ``{The Gravity
  Dual of a Density Matrix},''
  \href{http://dx.doi.org/10.1088/0264-9381/29/15/155009}{{\em
  Class.Quant.Grav.} {\bfseries 29} (2012) 155009},
\href{http://arxiv.org/abs/1204.1330}{{\ttfamily arXiv:1204.1330 [hep-th]}}.
%%CITATION = ARXIV:1204.1330;%%.

\bibitem{Bousso:2012sj}
R.~Bousso, S.~Leichenauer, and V.~Rosenhaus, ``{Light-sheets and AdS/CFT},''
  \href{http://dx.doi.org/10.1103/PhysRevD.86.046009}{{\em Phys.Rev.}
  {\bfseries D86} (2012) 046009},
\href{http://arxiv.org/abs/1203.6619}{{\ttfamily arXiv:1203.6619 [hep-th]}}.
%%CITATION = ARXIV:1203.6619;%%.

\bibitem{Emparan:1999gf}
R.~Emparan, ``{AdS / CFT duals of topological black holes and the entropy of
  zero energy states},''
  \href{http://dx.doi.org/10.1088/1126-6708/1999/06/036}{{\em JHEP} {\bfseries
  9906} (1999) 036},
\href{http://arxiv.org/abs/hep-th/9906040}{{\ttfamily arXiv:hep-th/9906040
  [hep-th]}}.
%%CITATION = HEP-TH/9906040;%%.

\bibitem{Czech:2012be}
B.~Czech, J.~L. Karczmarek, F.~Nogueira, and M.~Van~Raamsdonk, ``{Rindler
  Quantum Gravity},''
  \href{http://dx.doi.org/10.1088/0264-9381/29/23/235025}{{\em
  Class.Quant.Grav.} {\bfseries 29} (2012) 235025},
\href{http://arxiv.org/abs/1206.1323}{{\ttfamily arXiv:1206.1323 [hep-th]}}.
%%CITATION = ARXIV:1206.1323;%%.

\bibitem{Parikh:2012kg}
M.~Parikh and P.~Samantray, ``{Rindler-AdS/CFT},''
\href{http://arxiv.org/abs/1211.7370}{{\ttfamily arXiv:1211.7370 [hep-th]}}.
%%CITATION = ARXIV:1211.7370;%%.

\bibitem{Bousso:2012mh}
R.~Bousso, B.~Freivogel, S.~Leichenauer, V.~Rosenhaus, and C.~Zukowski, ``{Null
  Geodesics, Local CFT Operators and AdS/CFT for Subregions},''
  \href{http://dx.doi.org/10.1103/PhysRevD.88.064057}{{\em Phys.Rev.}
  {\bfseries D88} (2013) 064057},
\href{http://arxiv.org/abs/1209.4641}{{\ttfamily arXiv:1209.4641 [hep-th]}}.
%%CITATION = ARXIV:1209.4641;%%.

\bibitem{Ryu:2006bv}
S.~Ryu and T.~Takayanagi, ``{Holographic derivation of entanglement entropy
  from AdS/CFT},'' \href{http://dx.doi.org/10.1103/PhysRevLett.96.181602}{{\em
  Phys. Rev. Lett.} {\bfseries 96} (2006) 181602},
\href{http://arxiv.org/abs/hep-th/0603001}{{\ttfamily arXiv:hep-th/0603001}}.
%%CITATION = HEP-TH/0603001;%%.

\bibitem{Hubeny:2007aa}
V.~E. Hubeny, M.~Rangamani, and T.~Takayanagi, ``A covariant holographic
  entanglement entropy proposal,''
  \href{http://dx.doi.org/10.1088/1126-6708/2007/07/062}{{\em Journal of High
  Energy Physics} {\bfseries 2007} no.~07, (2007) 062},
  \href{http://arxiv.org/abs/0705.0016}{{\ttfamily arXiv:0705.0016 [hep-th]}}.

\bibitem{Lewkowycz:2013nqa}
A.~Lewkowycz and J.~Maldacena, ``{Generalized gravitational entropy},''
  \href{http://dx.doi.org/10.1007/JHEP08(2013)090}{{\em JHEP} {\bfseries 1308}
  (2013) 090},
\href{http://arxiv.org/abs/1304.4926}{{\ttfamily arXiv:1304.4926 [hep-th]}}.
%%CITATION = ARXIV:1304.4926;%%.

\bibitem{Faulkner:2013ana}
T.~Faulkner, A.~Lewkowycz, and J.~Maldacena, ``{Quantum corrections to
  holographic entanglement entropy},''
  \href{http://dx.doi.org/10.1007/JHEP11(2013)074}{{\em JHEP} {\bfseries 1311}
  (2013) 074},
\href{http://arxiv.org/abs/1307.2892}{{\ttfamily arXiv:1307.2892}}.
%%CITATION = ARXIV:1307.2892;%%.

\bibitem{Hubeny:2012wa}
V.~E. Hubeny and M.~Rangamani, ``{Causal Holographic Information},''
  \href{http://dx.doi.org/10.1007/JHEP06(2012)114}{{\em JHEP} {\bfseries 1206}
  (2012) 114},
\href{http://arxiv.org/abs/1204.1698}{{\ttfamily arXiv:1204.1698 [hep-th]}}.
%%CITATION = ARXIV:1204.1698;%%.

\bibitem{Casini:2009aa}
H.~Casini and M.~Huerta, ``Remarks on the entanglement entropy for disconnected
  regions,'' {\em Journal of High Energy Physics} {\bfseries 2009} no.~03,
  (2009) 048.

\bibitem{Hayden:2011ag}
P.~Hayden, M.~Headrick, and A.~Maloney, ``{Holographic Mutual Information is
  Monogamous},'' \href{http://arxiv.org/abs/1107.2940}{{\ttfamily
  arXiv:1107.2940 [hep-th]}}.
21 pages, 1 figure.
%%CITATION = ARXIV:1107.2940;%%.

\bibitem{Morrison:2012ab}
I.~A. Morrison and M.~M. Roberts, ``{Mutual information between thermo-field
  doubles and disconnected holographic boundaries},''
  \href{http://dx.doi.org/10.1007/JHEP07(2013)081}{{\em JHEP} {\bfseries 2013}
  no.~7, (2013) 1--31}, \href{http://arxiv.org/abs/1211.2887}{{\ttfamily
  arXiv:1211.2887 [hep-th]}}.

\bibitem{Wall:2012uf}
A.~C. Wall, ``{Maximin Surfaces, and the Strong Subadditivity of the Covariant
  Holographic Entanglement Entropy},''
\href{http://arxiv.org/abs/1211.3494}{{\ttfamily arXiv:1211.3494 [hep-th]}}.
%%CITATION = ARXIV:1211.3494;%%.

\bibitem{Hartman:2013qma}
T.~Hartman and J.~Maldacena, ``{Time Evolution of Entanglement Entropy from
  Black Hole Interiors},''
  \href{http://dx.doi.org/10.1007/JHEP05(2013)014}{{\em JHEP} {\bfseries 1305}
  (2013) 014},
\href{http://arxiv.org/abs/1303.1080}{{\ttfamily arXiv:1303.1080 [hep-th]}}.
%%CITATION = ARXIV:1303.1080;%%.

\bibitem{Blanco:2013joa}
D.~D. Blanco, H.~Casini, L.-Y. Hung, and R.~C. Myers, ``{Relative Entropy and
  Holography},''
\href{http://arxiv.org/abs/1305.3182}{{\ttfamily arXiv:1305.3182 [hep-th]}}.
%%CITATION = ARXIV:1305.3182;%%.

\bibitem{Kelly:2013aja}
W.~R. Kelly and A.~C. Wall, ``{Coarse-grained entropy and causal holographic
  information in AdS/CFT},''
\href{http://arxiv.org/abs/1309.3610}{{\ttfamily arXiv:1309.3610 [hep-th]}}.
%%CITATION = ARXIV:1309.3610;%%.

\bibitem{Kraus:2002iv}
P.~Kraus, H.~Ooguri, and S.~Shenker, ``Inside the horizon with ads/cft,''
  \href{http://dx.doi.org/10.1103/PhysRevD.67.124022}{{\em Phys. Rev. D}
  {\bfseries 67} (Jun, 2003) 124022},
  \href{http://arxiv.org/abs/hep-th/0212277}{{\ttfamily arXiv:hep-th/0212277
  [hep-th]}}.

\bibitem{Fidkowski:2004aa}
L.~Fidkowski, V.~Hubeny, M.~Kleban, and S.~Shenker, ``The black hole
  singularity in ads/cft,''
  \href{http://dx.doi.org/10.1088/1126-6708/2004/02/014}{{\em Journal of High
  Energy Physics} {\bfseries 2004} no.~02, (2004) 014},
  \href{http://arxiv.org/abs/hep-th/0306170}{{\ttfamily ArXiv:hep-th/0306170
  [hep-th]}}.

\bibitem{Hubeny:2006yu}
V.~E. Hubeny, H.~Liu, and M.~Rangamani, ``{Bulk-cone singularities \&
  signatures of horizon formation in AdS/CFT},''
  \href{http://dx.doi.org/10.1088/1126-6708/2007/01/009}{{\em Journal of High
  Energy Physics} {\bfseries 2007} no.~01, (2007) 009},
\href{http://arxiv.org/abs/hep-th/0610041}{{\ttfamily arXiv:hep-th/0610041}}.
%%CITATION = HEP-TH/0610041;%%.

\bibitem{Hubeny:2012ry}
V.~E. Hubeny, ``{Extremal surfaces as bulk probes in AdS/CFT},''
  \href{http://dx.doi.org/10.1007/JHEP07(2012)093}{{\em JHEP} {\bfseries 1207}
  (2012) 093},
\href{http://arxiv.org/abs/1203.1044}{{\ttfamily arXiv:1203.1044 [hep-th]}}.
%%CITATION = ARXIV:1203.1044;%%.

\bibitem{Bena:1999jv}
I.~Bena, ``{On the construction of local fields in the bulk of AdS(5) and other
  spaces},'' \href{http://dx.doi.org/10.1103/PhysRevD.62.066007}{{\em
  Phys.Rev.} {\bfseries D62} (2000) 066007},
\href{http://arxiv.org/abs/hep-th/9905186}{{\ttfamily arXiv:hep-th/9905186
  [hep-th]}}.
%%CITATION = HEP-TH/9905186;%%.

\bibitem{Hamilton:2005ju}
A.~Hamilton, D.~Kabat, G.~Lifschytz, and D.~A. Lowe, ``Local bulk operators in
  ads/cft correspondence: A boundary view of horizons and locality,''
  \href{http://dx.doi.org/10.1103/PhysRevD.73.086003}{{\em Phys. Rev. D}
  {\bfseries 73} no.~8, (Apr, 2006) 086003},
\href{http://arxiv.org/abs/hep-th/0506118}{{\ttfamily arXiv:hep-th/0506118}}.
%%CITATION = HEP-TH/0506118;%%.

\bibitem{Hamilton:2006az}
A.~Hamilton, D.~Kabat, G.~Lifschytz, and D.~A. Lowe, ``Holographic
  representation of local bulk operators,''
  \href{http://dx.doi.org/10.1103/PhysRevD.74.066009}{{\em Phys. Rev. D}
  {\bfseries 74} no.~6, (Sep, 2006) 066009},
\href{http://arxiv.org/abs/hep-th/0606141}{{\ttfamily arXiv:hep-th/0606141}}.
%%CITATION = HEP-TH/0606141;%%.

\bibitem{Hamilton:2006fh}
A.~Hamilton, D.~Kabat, G.~Lifschytz, and D.~A. Lowe, ``Local bulk operators in
  ads/cft correspondence: A holographic description of the black hole
  interior,'' \href{http://dx.doi.org/10.1103/PhysRevD.75.106001}{{\em Phys.
  Rev. D} {\bfseries 75} no.~10, (May, 2007) 106001},
\href{http://arxiv.org/abs/hep-th/0612053}{{\ttfamily arXiv:hep-th/0612053}}.
%%CITATION = HEP-TH/0612053;%%.

\bibitem{Lowe:2008ra}
D.~A. Lowe and S.~Roy, ``Holographic description of asymptotically $ads2$
  collapse geometries,''
  \href{http://dx.doi.org/10.1103/PhysRevD.78.124017}{{\em Phys. Rev. D}
  {\bfseries 78} no.~12, (Dec, 2008) 124017},
\href{http://arxiv.org/abs/0810.1750}{{\ttfamily arXiv:0810.1750 [hep-th]}}.
%%CITATION = 0810.1750;%%.

\bibitem{Kabat:2011rz}
D.~Kabat, G.~Lifschytz, and D.~A. Lowe, ``{Constructing local bulk observables
  in interacting AdS/CFT},''
  \href{http://dx.doi.org/10.1103/PhysRevD.83.106009}{{\em Phys. Rev. D}
  {\bfseries 83} no.~10, (May, 2011) 106009},
\href{http://arxiv.org/abs/1102.2910}{{\ttfamily arXiv:1102.2910 [hep-th]}}.
%%CITATION = 1102.2910;%%.

\bibitem{Heemskerk:2012mq}
I.~Heemskerk, ``{Construction of Bulk Fields with Gauge Redundancy},''
\href{http://arxiv.org/abs/1201.3666}{{\ttfamily arXiv:1201.3666 [hep-th]}}.
%%CITATION = ARXIV:1201.3666;%%.

\bibitem{Kabat:2012hp}
D.~Kabat, G.~Lifschytz, S.~Roy, and D.~Sarkar, ``{Holographic representation of
  bulk fields with spin in AdS/CFT},''
\href{http://arxiv.org/abs/1204.0126}{{\ttfamily arXiv:1204.0126 [hep-th]}}.
%%CITATION = ARXIV:1204.0126;%%.

\bibitem{Kabat:2012av}
D.~Kabat and G.~Lifschytz, ``{CFT representation of interacting bulk gauge
  fields in AdS},''
\href{http://arxiv.org/abs/1212.3788}{{\ttfamily arXiv:1212.3788 [hep-th]}}.
%%CITATION = ARXIV:1212.3788;%%.

\bibitem{Leichenauer:2013kaa}
S.~Leichenauer and V.~Rosenhaus, ``{AdS black holes, the bulk-boundary
  dictionary, and smearing functions},''
  \href{http://dx.doi.org/10.1103/PhysRevD.88.026003}{{\em Phys.Rev.}
  {\bfseries D88} (2013) 026003},
\href{http://arxiv.org/abs/1304.6821}{{\ttfamily arXiv:1304.6821 [hep-th]}}.
%%CITATION = ARXIV:1304.6821;%%.

\bibitem{Breitenlohner:1982bm}
P.~Breitenlohner and D.~Z. Freedman, ``{Positive Energy in anti-De Sitter
  Backgrounds and Gauged Extended Supergravity},''
\href{http://dx.doi.org/10.1016/0370-2693(82)90643-8}{{\em Phys.Lett.}
  {\bfseries B115} (1982) 197}.
%%CITATION = PHLTA,B115,197;%%.

\bibitem{Breitenlohner:1982jf}
P.~Breitenlohner and D.~Z. Freedman, ``{Stability in Gauged Extended
  Supergravity},''
\href{http://dx.doi.org/10.1016/0003-4916(82)90116-6}{{\em Annals Phys.}
  {\bfseries 144} (1982) 249}.
%%CITATION = APNYA,144,249;%%.

\bibitem{Ishibashi:2004wx}
A.~Ishibashi and R.~M. Wald, ``{Dynamics in nonglobally hyperbolic static
  space-times. 3. Anti-de Sitter space-time},''
  \href{http://dx.doi.org/10.1088/0264-9381/21/12/012}{{\em Class.Quant.Grav.}
  {\bfseries 21} (2004) 2981--3014},
\href{http://arxiv.org/abs/hep-th/0402184}{{\ttfamily arXiv:hep-th/0402184
  [hep-th]}}.
%%CITATION = HEP-TH/0402184;%%.

\bibitem{Friedlander:1975aa}
F.~G. Friedlander, {\em {The wave equation on a curved space-time}}.
\newblock Cambridge Monographs on Mathematical Physics. Cambridge University
  Press, 1975.
\newblock 296 p.

\bibitem{Hollands:2001fk}
S.~Hollands and R.~M. Wald, ``Local wick polynomials and time ordered products
  of quantum fields in curved spacetime,''
  \href{http://dx.doi.org/10.1007/s002200100540}{{\em Comm. Math. Phys.}
  {\bfseries 223} no.~2, (2001) 289--326},
  \href{http://arxiv.org/abs/gr-qc/0103074}{{\ttfamily gr-qc/0103074}}.

\bibitem{Hollands:2002ux}
S.~Hollands and R.~M. Wald, ``{On the renormalization group in curved
  space-time},'' {\em Commun.Math.Phys.} {\bfseries 237} (2003) 123--160,
\href{http://arxiv.org/abs/gr-qc/0209029}{{\ttfamily arXiv:gr-qc/0209029
  [gr-qc]}}.
%%CITATION = GR-QC/0209029;%%.

\bibitem{Hollands:2004yh}
S.~Hollands and R.~M. Wald, ``{Conservation of the stress tensor in interacting
  quantum field theory in curved spacetimes},''
  \href{http://dx.doi.org/10.1142/S0129055X05002340}{{\em Rev.Math.Phys.}
  {\bfseries 17} (2005) 227--312},
\href{http://arxiv.org/abs/gr-qc/0404074}{{\ttfamily arXiv:gr-qc/0404074
  [gr-qc]}}.
%%CITATION = GR-QC/0404074;%%.

\bibitem{Brunetti:1995rf}
R.~Brunetti, K.~Fredenhagen, and M.~Kohler, ``{The Microlocal spectrum
  condition and Wick polynomials of free fields on curved space-times},''
  \href{http://dx.doi.org/10.1007/BF02099626}{{\em Comm. Math. Phys.}
  {\bfseries 180} (1996) 633--652},
\href{http://arxiv.org/abs/gr-qc/9510056}{{\ttfamily arXiv:gr-qc/9510056
  [gr-qc]}}.
%%CITATION = GR-QC/9510056;%%.

\bibitem{Burgess:1985aa}
C.~P. Burgess and C.~A. L¸tken, ``Propagators and effective potentials in
  anti-de sitter space,'' \href{http://dx.doi.org/DOI:
  10.1016/0370-2693(85)91415-7}{{\em Physics Letters B} {\bfseries 153} no.~3,
  (1985) 137 -- 141}.

\bibitem{Streater:1989vi}
R.~F. Streater and A.~S. Wightman, {\em {PCT, spin and statistics, and all
  that}}.
\newblock Advanced book classics. Addison-Wesley, Redwood City, USA, 1989.
\newblock 207 p.

\bibitem{Strohmaier:2002aa}
A.~Strohmaier, R.~Verch, and M.~Wollenberg, ``Microlocal analysis of quantum
  fields on curved space--times: Analytic wave front sets and reeh--schlieder
  theorems,'' \href{http://dx.doi.org/10.1063/1.1506381}{{\em J. Math. Phys.}
  {\bfseries 43} no.~11, (2002) 5514--5530},
  \href{http://arxiv.org/abs/math-ph/0202003}{{\ttfamily math-ph/0202003
  [math-ph]}}.

\bibitem{Heemskerk:2012mn}
I.~Heemskerk, D.~Marolf, and J.~Polchinski, ``{Bulk and Transhorizon
  Measurements in AdS/CFT},''
\href{http://arxiv.org/abs/1201.3664}{{\ttfamily arXiv:1201.3664 [hep-th]}}.
%%CITATION = ARXIV:1201.3664;%%.

\bibitem{Casini:2011kv}
H.~Casini, M.~Huerta, and R.~C. Myers, ``{Towards a derivation of holographic
  entanglement entropy},''
  \href{http://dx.doi.org/10.1007/JHEP05(2011)036}{{\em JHEP} {\bfseries 1105}
  (2011) 036},
\href{http://arxiv.org/abs/1102.0440}{{\ttfamily arXiv:1102.0440 [hep-th]}}.
%%CITATION = ARXIV:1102.0440;%%.

\bibitem{Unruh:1984aa}
W.~G. Unruh and R.~M. Wald, ``What happens when an accelerating observer
  detects a rindler particle,''
  \href{http://dx.doi.org/10.1103/PhysRevD.29.1047}{{\em Phys. Rev. D}
  {\bfseries 29} (Mar, 1984) 1047--1056}.

\bibitem{Birrell:1982ix}
N.~D. Birrell and P.~C.~W. Davies, {\em {Quantum fields in curved space}}.
\newblock Cambridge University Press, Cambridge, UK, 1982.
\newblock 340p.

\bibitem{Aminneborg:1997aa}
S.~{\AA}minneborg, I.~Bengtsson, D.~Brill, S.~Holst, and P.~Peld{\'a}n, ``Black
  holes and wormholes in $2+1$ dimensions,'' {\em Classical and Quantum
  Gravity} {\bfseries 15} no.~3, (1998) 627.

\bibitem{Krasnov:2000zq}
K.~Krasnov, ``Holography and riemann surfaces,'' {\em Adv.Theor.Math.Phys.}
  {\bfseries 4} (2000) 929--979,
  \href{http://arxiv.org/abs/hep-th/0005106}{{\ttfamily arXiv:hep-th/0005106
  [hep-th]}}.

\bibitem{Krasnov:2003aa}
K.~Krasnov, ``Black-hole thermodynamics and riemann surfaces,''
  \href{http://dx.doi.org/10.1088/0264-9381/20/11/319}{{\em Classical and
  Quantum Gravity} {\bfseries 20} no.~11, (2003) 2235},
  \href{http://arxiv.org/abs/gr-qc/0302073}{{\ttfamily arXiv:gr-qc/0302073
  [gr-qc]}}.

\bibitem{Skenderis:2009ju}
K.~Skenderis and B.~C. van Rees, ``Holography and wormholes in 2+1
  dimensions,'' \href{http://dx.doi.org/10.1007/s00220-010-1163-z}{{\em
  Commun.Math.Phys.} {\bfseries 301} (2011) 583--626},
  \href{http://arxiv.org/abs/0912.2090}{{\ttfamily arXiv:0912.2090 [hep-th]}}.

\bibitem{Shenker:2013yza}
S.~H. Shenker and D.~Stanford, ``{Multiple Shocks},''
\href{http://arxiv.org/abs/1312.3296}{{\ttfamily arXiv:1312.3296 [hep-th]}}.
%%CITATION = ARXIV:1312.3296;%%.

\bibitem{Papadodimas:2012aq}
K.~Papadodimas and S.~Raju, ``{An Infalling Observer in AdS/CFT},''
\href{http://arxiv.org/abs/1211.6767}{{\ttfamily arXiv:1211.6767 [hep-th]}}.
%%CITATION = ARXIV:1211.6767;%%.

\bibitem{Verlinde:2013qya}
E.~Verlinde and H.~Verlinde, ``{Behind the Horizon in AdS/CFT},''
\href{http://arxiv.org/abs/1311.1137}{{\ttfamily arXiv:1311.1137 [hep-th]}}.
%%CITATION = ARXIV:1311.1137;%%.

\bibitem{Avery:2013bea}
S.~G. Avery and B.~D. Chowdhury, ``{No Holography for Eternal AdS Black
  Holes},''
\href{http://arxiv.org/abs/1312.3346}{{\ttfamily arXiv:1312.3346 [hep-th]}}.
%%CITATION = ARXIV:1312.3346;%%.

\bibitem{Balasubramanian:2006aa}
V.~Balasubramanian, B.~Czech, K.~Larjo, and J.~Sim\'on, ``Integrability vs.
  information loss: a simple example,'' {\em Journal of High Energy Physics}
  {\bfseries 2006} no.~11, (2006) 001.

\bibitem{Balasubramanian:2006ab}
V.~Balasubramanian, D.~Marolf, and M.~Rozali, ``Information recovery from black
  holes,'' \href{http://dx.doi.org/10.1007/s10714-006-0344-8}{{\em General
  Relativity and Gravitation} {\bfseries 38} (2006) 1529--1536}.

\bibitem{Marolf:2008mf}
D.~Marolf, ``{Unitarity and Holography in Gravitational Physics},''
  \href{http://dx.doi.org/10.1103/PhysRevD.79.044010}{{\em Phys.Rev.}
  {\bfseries D79} (2009) 044010},
\href{http://arxiv.org/abs/0808.2842}{{\ttfamily arXiv:0808.2842 [gr-qc]}}.
%%CITATION = ARXIV:0808.2842;%%.

\bibitem{Marolf:2008mg}
D.~Marolf, ``{Holographic Thought Experiments},''
  \href{http://dx.doi.org/10.1103/PhysRevD.79.024029}{{\em Phys.Rev.}
  {\bfseries D79} (2009) 024029},
\href{http://arxiv.org/abs/0808.2845}{{\ttfamily arXiv:0808.2845 [gr-qc]}}.
%%CITATION = ARXIV:0808.2845;%%.

\bibitem{Marolf:2013iba}
D.~Marolf, ``{Holography without strings?},''
\href{http://arxiv.org/abs/1308.1977}{{\ttfamily arXiv:1308.1977 [hep-th]}}.
%%CITATION = ARXIV:1308.1977;%%.

\bibitem{Hormander:1990aa}
L.~H\"ormander, {\em {The Analysis of Linear Partial Differential Operators I:
  Distribution Theory and Fourier Analysis}}.
\newblock Springer-Verlag, Berlin / Heidelberg, 2nd~ed., 1990.
\newblock 440 p.

\bibitem{Radzikowski:1996aa}
M.~J. Radzikowski, ``Micro-local approach to the hadamard condition in quantum
  field theory on curved space-time,''
  \href{http://dx.doi.org/10.1007/BF02100096}{{\em Comm. Math. Phys.}
  {\bfseries 179} (1996) 529--553}.

\bibitem{Verch:1999aa}
R.~Verch, ``Wavefront sets in algebraic quantum field theory,''
  \href{http://dx.doi.org/10.1007/s002200050680}{{\em Comm. Math. Phys.}
  {\bfseries 205} (1999) 337--367}.

\end{thebibliography}\endgroup

%%%%%%%%%%%%%%%%%%%%%%%%%%%%%%%%%%%%%%%%%%%%%%%%%%%%%%%%%%%%%%%%%
\end{document}